\documentclass[a4paper,12pt]{article}

\usepackage{jheppub} 

\usepackage[T1]{fontenc} 
\usepackage{bm,latexsym,amsmath,amssymb,amsfonts,mathtools,mathrsfs}
\usepackage{comment}
\usepackage{ulem}
\usepackage{color}

\usepackage{bm}
\usepackage{hyperref}
\usepackage{framed}
\usepackage{subfigure}
\usepackage{hyperref}   
\usepackage{graphicx}

\expandafter\let\csname equation*\endcsname\relax 
\expandafter\let\csname endequation*\endcsname\relax 
\usepackage{amsmath}
\usepackage{color}

\def\red{}

\usepackage{hyperref}
\hypersetup{
    colorlinks=true,
    linkcolor=blue,
    filecolor=magenta,
    urlcolor=blue,
}
\newcommand{\ct}{{(\rm T)}}

\newcommand{\be}{\begin{equation}}
\newcommand{\ee}{\end{equation}}
\newcommand{\beal}{\begin{aligned}}
\newcommand{\eeal}{\end{aligned}}
\newcommand\bea {\begin{eqnarray}}
\newcommand\eea {\end{eqnarray}}
\newcommand{\bec}{\begin{cases}}
\newcommand{\eec}{\end{cases}}

\begin{document}

\title{Black hole evaporation in de Sitter space}

\author[a,b]{Ruth Gregory}
\author[c]{Ian G. Moss}
\author[b]{Naritaka Oshita}
\author[d]{Sam Patrick}

\emailAdd{ruth.gregory@kcl.ac.uk}
\emailAdd{ian.moss@newcastle.ac.uk}
\emailAdd{naritaka.oshita@gmail.com}
\emailAdd{sampatrick31@googlemail.com}

\affiliation[a]{Department of Physics, King's College London,
The Strand, London WC2R 2LS, UK}
\affiliation[b]{Perimeter Institute, 31 Caroline Street North, Waterloo, 
ON, N2L 2Y5, Canada}
\affiliation[c]{School of Mathematics, Statistics and Physics, Newcastle University, 
Newcastle Upon Tyne, NE1 7RU, UK}
\affiliation[d]{Department of Physics and Astronomy, The University of British
Columbia, Vancouver, Canada, V6T 1Z1}

\date{\today}

\abstract{
We investigate the evaporation process of a Kerr-de Sitter black hole with the Unruh-Hawking-like 
vacuum state, which is a realistic vacuum state modelling the evaporation process of a black hole 
originating from gravitational collapse. We also compute the greybody factors for gravitons, photons, 
and conformal-coupling massless scalar particles by using the analytic solutions of the Teukolsky equation in the Kerr-de 
Sitter background. It turns out that the cosmological constant quenches the amplification factor
and it approaches to zero towards the critical point where the Nariai and extremal limits merge together. We confirm that even 
near the critical point, the superradiance of gravitons is more significant than that of photons and scalar particles.
Angular momentum is carried out by particles several times faster than the mass energy decreases.
This means that a Kerr-de Sitter black hole rapidly spins down to a nearly Schwarzschild-de Sitter black hole before it completely evaporates.
We also compute the time evolution of the Bekenstein-Hawking entropy. The total entropy of the Kerr-de Sitter black hole and cosmological horizon
increases with time, which is consistent with the generalized second law of thermodynamics.
}
\maketitle

\section{Introduction}

The theory of black hole evaporation stands as something of a milestone in theoretical
physics.  In the context of quantum field theory on a curved spacetime background,
black holes radiate thermal radiation, modified by scattering effects
from the spacetime geometry. The backreaction of the radiation flux
causes the black hole to evaporate, and black holes
are expected to lose all their mass and angular momentum.
In this paper we address the fate of a black hole in de Sitter space, where
the black hole and the surrounding spacetime have potentially different
temperatures. We focus on the situation where the black hole 
and the de Sitter space both have a finite past history,
such as might arise in the early universe with black holes that form from
gravitational collapse.

The evolution of mass and angular momentum for black holes in empty space
was first discussed in the pioneering papers of Hawking 
\cite{Hawking:1974rv,Hawking:1974sw},
where it was proposed that the black hole loses mass at an increasing rate until nothing is left.
The numerical details were supplied by Page \cite{Page:1976df} and extended to
rotating holes a little later \cite{Page:1976ki}. Page found that the black hole
always loses angular momentum more rapidly than it loses mass. The backreaction of
radiation emitted by the black hole can also be studied directly via the quantum
expectation value of the stress-energy tensor \cite{PhysRevD.21.2185},
and there has been recent progress calculating the full quantum stress
tensor \cite{Levi:2015eea,Levi:2016paz,Levi:2016exv}. 

The evolution of mass and angular momentum are related to the fluxes of energy and angular momentum
from quantum fields around the black hole. These fluxes are expectation values, or ensemble averaged
expectation values, of a quantum stress tensor $\hat T_{\mu\nu}$ \cite{PhysRevD.21.2185}. 
The choice of quantum states plays 
a crucial role in evaluating these expectation values. We illustrate this by reviewing what happens for
an eternal black hole background in an asymptotically flat spacetime.

\begin{figure} 
\centering
\includegraphics[width=0.3\linewidth]{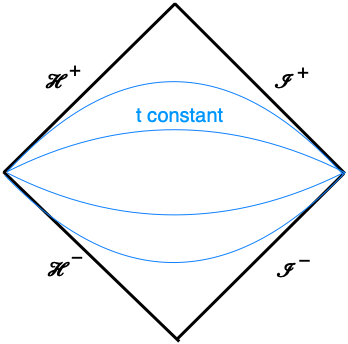}\hspace{1cm}
\includegraphics[width=0.3\linewidth]{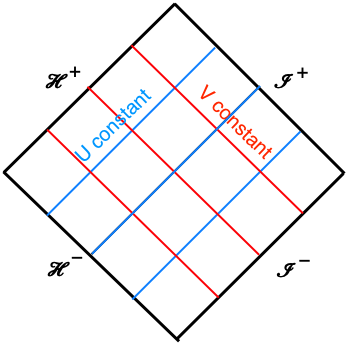}
\caption{Penrose diagram for the exterior of a black hole in asymptotically flat spacetime.
} \label{fig:hole1}
\end{figure}

Three quantum vacuum states play significant roles \cite{PhysRevD.14.870}. 
Each is defined with respect to sets
of positive frequency wave modes on the background spacetime:

\begin{itemize}
\item the Boulware vacuum $|B\rangle$, defined by taking modes which are positive 
frequency with respect to the Killing vector $\partial_t$, with respect to which the exterior 
region is static,
\item the Unruh-Hawking vacuum $|U\rangle$, defined by taking modes that 
are incoming from $\mathscr{I}^-$ 
to be positive frequency with respect to $\partial_t$, while those that 
emanate from the past  horizon are taken to be positive frequency with respect to $\partial_U$, 
where $U$ is the canonical affine parameter on the past horizon,
\item the Hartle-Hawking vacuum $|H\rangle$, defined by taking incoming modes to be positive 
frequency with respect to $\partial_V$, where $V$ is the canonical affine parameter on the 
future horizon, and outgoing modes to be positive frequency 
with respect to $\partial_U$.
\end{itemize}

The vacuum state is said to be regular on the horizon if the Feynman two-point function for that
vacuum state is regular there. In practice, this means that the first derivative
of the Feynman Green function with respect to the Kruskal coordinates is finite.
The Boulware vacuum state is not regular on either the past or future black hole horizons,
which nevertheless makes it suitable for the space outside a neutron star, for example, where the spacetime is locally
isometric to the black hole metric but neither horizon is present.

The Unruh-Hawking vacuum state is used for a black hole formed from 
gravitational collapse. 
In reality, the collapse would result in a highly excited state.
However, the no hair theorems imply that the initial excitations would rapidly decay away
and the state would be expected to evolve to the vacuum state over a short period of time. 
Despite being a vacuum state, the
particle flux from the hole is non-vanishing, and has a thermal spectrum
at the Hawking temperature modified by particle scattering from the spacetime geometry,
and super-radiance contributions if the black hole is rotating. 

The affine parameter $U$, shown
on the right in figure \ref{fig:hole1}, vanishes at the future black hole horizon. The 
Unruh-Hawking vacuum state is regular on $U=0$.
The state is not regular on the past black hole event horizon, where the derivative
of the Green's functions with respect to $V$ diverges. 

\begin{figure} 
\centering
\includegraphics[width=0.2\linewidth]{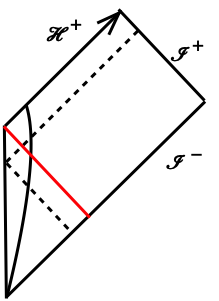}
\caption{Penrose diagram for a collapsing star.
} \label{fig:hole2}
\end{figure}

The past horizon is not present in the typical collapse situation shown in 
figure \ref{fig:hole2}. Instead, there is a null surface spanned by infalling null rays converging on the
point where the future event horizon forms. We can think of this surface as a
virtual past horizon. In the geometrical optics limit, wave vectors follow the paths of null
geodesics. Hence, high-frequency infalling waves with wave vector $\partial_v$, that pass through the
the collapsing star, emerge as outgoing waves with wave vector $\partial_V$.
This converts a fraction of the positive frequency $\partial_t$ modes into
outgoing positive frequency $\partial_V$ modes.

The Hartle-Hawking vacuum is used for the `black hole in box', to
represent a thermodynamical ensemble at the Hawking temperature. The vacuum
state is regular on both the past and future black hole horizons, and particle fluxes are
in perfect balance. However, gravitational backreaction from the Hawking radiation
is not consistent with having an asymptotically flat spacetime, hence the need
for a reflecting barrier around the hole to cut off the radiation.

\begin{figure} 
\centering
\includegraphics[width=0.3\linewidth]{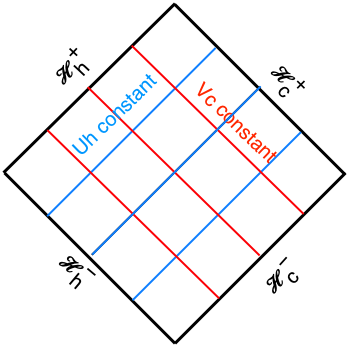}\hspace{2cm}
\includegraphics[width=0.3\linewidth]{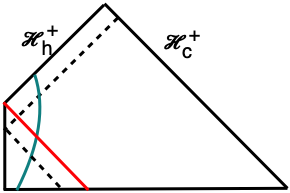}
\caption{Penrose diagrams for the exterior region of an eternal black hole in de Sitter space and
a collapsing region in de Sitter space starting from a spacelike surface. The affine parameter
$U_h=-e^{-\kappa_h u}$ is zero on the future black hole horizon ${\cal H}_h^+$,
and $V_c=-e^{-\kappa_c v}$ is zero on the future cosmological horizon ${\cal H}_c^+$.
} \label{fig:sds}
\end{figure}

We turn now to the black hole in de Sitter space. The cosmological horizon of de Sitter space emits
particles with temperature $T_c$ \cite{PhysRevD.15.2738} which is typically different from the black hole
temperature $T_h$. There are two particle fluxes in competition, an incoming flux
from the cosmological horizon and an outgoing flux from the black hole. To represent this situation
we generalise the Unruh-Hawking vacuum state as follows:

\begin{quote}\parindent=1cm
The vacuum state $|V\rangle$  is defined by taking modes that are incoming from the past cosmological horizon
to be positive frequency with respect to $\partial_{V_c}$, while those that emanate from the past 
black hole horizon are taken to be positive frequency with respect to $\partial_{U_h}$, 
the canonical affine parameter on the past horizon.
\end{quote}

A similar vacuum state has been used to analyse particle production from non-rotating
black holes in de Sitter space in \cite{Qiu:2019qgp,Anderson:2020dim}. Note that the state is not the `Unruh'
state used for de Sitter space in \cite{Aalsma:2019rpt,Gong:2020mbn}, where the
authors use ingoing modes for $\partial_{U_c}$ and outgoing for $\partial_t$.

The same argument given in Ref \cite{PhysRevD.14.870} 
implies that this vacuum state is regular on the future 
black hole horizon, where the parameter $U_h$ vanishes,
and on the future cosmological horizon, where the parameter $V_c$ vanishes.
We cannot say for sure that this is the unique vacuum state with this property, 
in the sense that there is a unique vacuum state which is regular
on the future and past black hole horizons in asymptotically flat space \cite{KAY199149}.

Consider next how the state could emerge physically.
Globally, de Sitter space represents a universe that collapses from the infinite past, bounces
and then expands to the infinite future. In a big bang cosmology, de Sitter space arises locally
as part of the inflationary universe scenario. The collapse phase is absent,
as is the past cosmological horizon. However, the initial quantum state of the inflationary
universe  is usually taken to be the Bunch-Davies vacuum state \cite{Bunch1978}, which is positive frequency
with respect to the de Sitter killing vectors $\partial_{U_c}$ and $\partial_{V_c}$. 
Now suppose a black hole forms. In the geometrical optics limit, ingoing waves with wave vector
$\partial_{V_c}$ that pass through the 
collapsing body are converted into outgoing waves with wave vector $\partial_{U_h}$,
as shown in the Penrose diagram on the right of figure \ref{fig:sds}. A fraction
of the incoming modes should therefore become positive frequency $\partial_{U_h}$ modes.

As in the asymptotically flat case, we expect the excited quantum state to settle down to
the vacuum state at late (advanced) time. This occurs in the classical theory, 
where it has been
shown that the no hair theorem extends to black holes in de Sitter space 
\cite{Mellor:1989ac,Chambers:1994sz}. 
The decay rate of perturbations is exponential at late cosmological time \cite{Chambers:1994sz}. 
The outgoing waves with wave vector
$\partial_{U_c}$ that are present in the initial state should also decay away, leaving
the positive frequency $\partial_{U_h}$ modes.

Given the vacuum state $|V\rangle$, the evolution of the mass $M$ and angular momentum 
$J$ of the black hole in de Sitter space with cosmological constant $3/\ell^2$ are
governed by the energy and angular momentum fluxes $\langle V|\hat T_{rt}|V\rangle$ and
$\langle V|\hat T_{r\varphi}|V\rangle$. These will be calculated in section \ref{sec:flux} from the appropriate mode
decompositions of scalar quantum fields. We will arrive at pleasingly compact results
for a black hole in de Sitter space,
\begin{align}
{dM\over dt}&=-\sum_{lm_z}\int_0^\infty {d\omega\over 2\pi}\left(\omega-{J\over M\ell^2}m_z\right)
\left(1-|A_{\omega lm_z}|^2\right)\frac12\left\{{\rm coth}{\beta_h\omega_h\over 2}
-{\rm coth}{\beta_c\omega_c\over 2}\right\},\\
{dJ\over dt}&=-\sum_{lm_z}\int_0^\infty {d\omega\over 2\pi}\,m_z
\left(1-|A_{\omega lm_z}|^2\right)\frac12\left\{{\rm coth}{\beta_h\omega_h\over 2}
-{\rm coth}{\beta_c\omega_c\over 2}\right\},
\end{align}
where $\beta_i$ are inverse temperatures at the black hole ($i=h$) and cosmological ($i=c$) horizons, 
respectively. The horizon frequencies $\omega_i \equiv \omega-m_z\Omega_i$  in terms of the 
horizon rotation $\Omega_i$, and the factors $A_{\omega lm_z}$ are reflection coefficients for 
waves transiting the spacetime geometry. Finally, 
$l$ and $m_{z}$ are the angular and 
azimuthal modes, respectively. One should be careful not to 
confuse the angular mode $l$ with the de Sitter radius $\ell$.

Special cases include:
\begin{itemize}
\item The Boulware vacuum $\beta_h\to\infty$, $\beta_c\to\infty$,
\begin{equation}
{dM\over dt}=\sum_{lm_{z}}\int_{\omega_c\omega_h<0} {d\omega\over 4\pi} |\omega|
\left(1-|A_{\omega lm_{z}}|^2\right),
\end{equation}
in agreement with Unruh, \cite{PhysRevD.10.3194}.
\item The Unruh-Hawking vacuum (non-rotating case) $\beta_c\to\infty$, $\omega_h=\omega_c=\omega$,
\begin{equation}
{dM\over dt}=-\sum_{lm_{z}}\int_0^\infty {d\omega\over 2\pi}\omega
\left(1-|A_{\omega lm_{z}}|^2\right){1\over e^{\beta_h\omega}-1},
\end{equation}
as originally derived by Hawking \cite{Hawking:1974sw}.
\item The Unruh-Hawking vacuum (rotating case) $\beta_c\to\infty$, $\omega_c=\omega$
\begin{equation}
{dM\over dt}=-\sum_{lm_{z}}\int_0^\infty {d\omega\over 2\pi}\omega
\left(1-|A_{\omega lm_{z}}|^2\right){1\over e^{\beta_h\omega_h}-1}.
\end{equation}
which was also derived by Hawking \cite{Hawking:1974sw}.
\end{itemize}
Note that the fluxes diverge in the $M\to 0$ or $\Lambda\to\infty$ limits, so the formulae don't apply to
the case $M=0$.

The evolution of mass and angular momentum is described in detail in 
\S \ref{evaporation}.
The most difficult part of the calculation is evaluating the reflection and transmissions factors.
We have evaluated these for the combinations of $(l,m_z)$ that give the largest contributions to the
fluxes. Results are given for conformally coupled scalar, electromagnetic, and gravitational fields.

We find that a rotating black hole with cosmological constant spins down to a nearly non-spinning black hole before 
most of its mass has been carried away by the Hawking radiation.
The loss of angular momentum is caused mostly by superradiance, but it turns out that the 
superradiance is weakened by a large value of cosmological constant.
We will show that the entropy of black hole horizon decreases while that of cosmological horizon increases with time.
The total entropy increases, which is consistent with the generalized second law of thermodynamics.

Throughout this paper we use Planck units with $G=c=\hbar=1$.

\section{Modes}

\begin{figure} 
\centering
\includegraphics[width=0.4\linewidth]{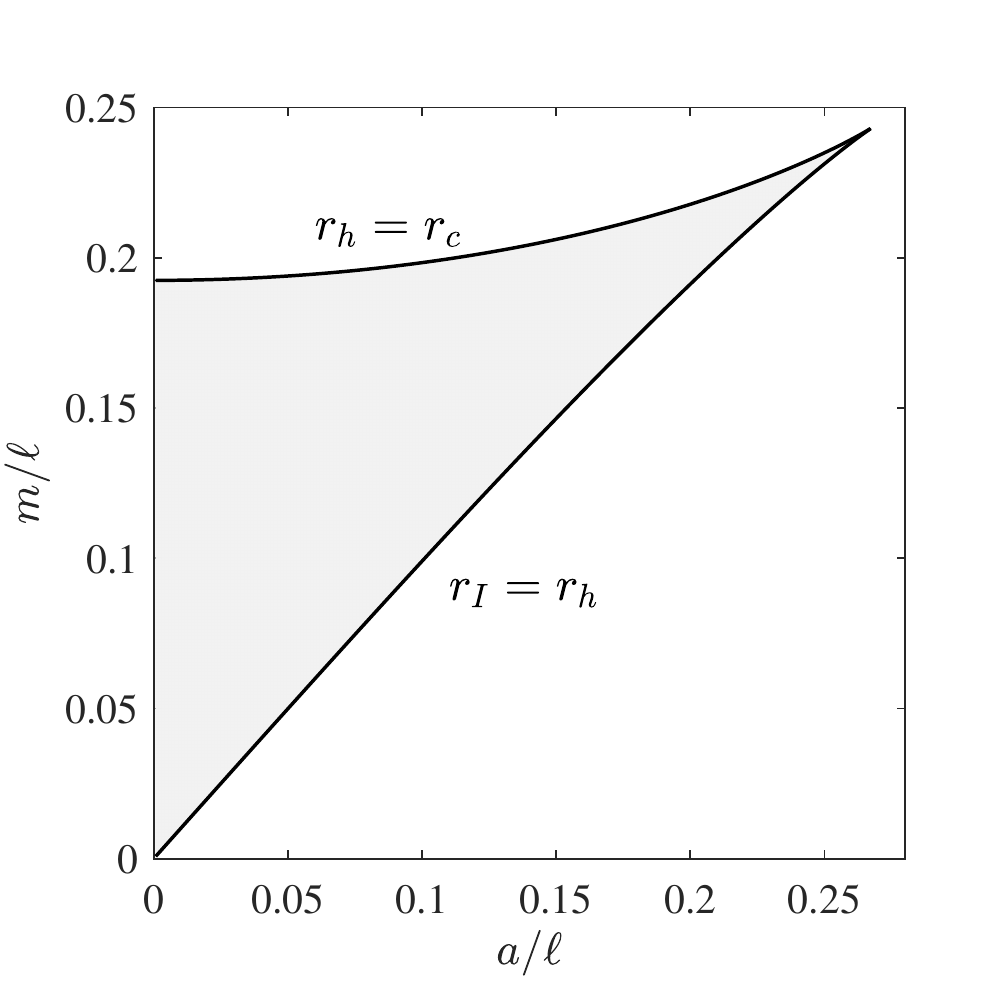}
\caption{Parameter space for the Kerr de Sitter metric. The extremal limit $r_I=r_h$ 
forms the right boundary of the plot whereas the Nariai limit $r_c=r_h$ forms the upper boundary.
Outside of the shaded region, one encounters naked singularities.
} \label{fig:params}
\end{figure}

We start with a few basic definitions. The rotating black hole in de Sitter space has metric
\begin{equation}
ds^2=-{\Delta_r\over\rho^2}\omega_t^2
+{\Delta_\theta\over\rho^2}\sin^2 \theta \, \omega_\varphi^2
+{\rho^2\over\Delta_r}dr^2+{\rho^2\over\Delta_\theta}d\theta^2,
\end{equation}
with the functions 
\begin{align}
\Delta_r&=(r^2+a^2)(1-r^2/\ell^2)-2mr,\\
\Delta_\theta&=1+{a^2\over \ell^2}\cos^2\theta,\\
\rho&=(r^2+a^2\cos^2\theta)^{1/2}.
\end{align}
The horizons of the spacetime are determined as the locations where $\Delta_r=0$ and are denoted $
r_I$ (inner horizon), $r_h$ (black hole horizon) and $r_c$ (cosmological horizon).
The basis forms are,
\begin{align}
\omega_t&=dt-a\Xi^{-1}\sin^2\theta\,d\varphi,\\
\omega_\varphi&=\Xi^{-1}(a^2+r^2)d\varphi-a dt,
\end{align}
where $\Xi=1+a^2/\ell^2$.
The space of metric parameters $(m,a,\ell)$ is effectively two dimensional and is illustrated in Fig. \ref{fig:params}.

The mass $M$ and the angular momentum $J$ can be defined using
analytic continuation from anti-de Sitter space \cite{Henneaux:1985tv},
and was studied in detail in the thermodynamical context in \cite{Dolan:2013ft}
(see also \cite{Dolan:2018hpl}). 
In terms of the mass parameter $m$ and the
rotation parameter $a$,
\begin{equation}
M={m\over\Xi^2},\qquad J={am\over \Xi^2}.
\end{equation}

We take a non-minimally coupled scalar field,  $\phi$, on the black hole background
with field equation
\begin{equation}
-\nabla_\mu\nabla^\mu\phi+\xi {\cal R} \phi=0,
\end{equation}
where ${\cal R}$ is the Ricci scalar and $\xi$ is the non-minimal coupling constant. 
Solutions are expanded in separable modes \cite{Chambers:1994ap}
\begin{equation}
\phi=\sum_{lm_z}\int{d\omega\over 2\pi}{R_{\omega lm_z}(r)\over (r^2+a^2)^{1/2}}S_{\omega lm_z}(\theta)
e^{im_z\varphi}e^{-i\omega t},\label{expand}
\end{equation}
where the $S_{\omega lm_z}(\theta)$'s are the angular eigenfunctions
\cite{Tachizawa:1992ue}. We introduce a tortoise coordinate defined by
\begin{equation}
dr^* \equiv {r^2+a^2\over \Delta_r}dr,
\end{equation}
The radial equation is then of the form
\begin{equation}
-\partial_{r^*}^2 R_{\omega lm_{z}}+V_{\omega l m_{z}}(r)R_{\omega lm_{z}}=0
\end{equation}
with the potential
\begin{equation}
V_{\omega l m_{z}} = - (\omega- m_{z} \Omega)^{2} + 
\frac{\Delta_{r} }{(r^{2}+a^{2})^{2}} \left\{\lambda+12\xi{r^2\over \ell^2}+ 
(r^{2}+a^{2})^{1/2}\left( \frac{r \Delta_{r}}{(r^{2}+a^{2})^{3/2}} \right)'\right\},
\end{equation}
where $\lambda$ is angular eigenvalue (that depends on $l$, $m_z$ and $\omega$), and
\begin{equation} \label{ang_vel}
\Omega \equiv \Xi{a\over r^2+a^2}.
\end{equation}
The potential has the limit of $V\to -(\omega-m_z\Omega)^2$ on the horizons.
There are ingoing or outgoing modes described here by arrows,
\begin{equation}\label{modes}
{\overrightarrow R}\to\sqrt{\Xi\over|2\omega_h|}
\begin{cases}
e^{i\omega_h r^*}+{\overrightarrow A} e^{-i\omega_h r^*}&r\to r_h\\
{\overrightarrow B} e^{i\omega_c r^*}&r\to r_c,
\end{cases}
\end{equation}
\begin{equation}
{\overleftarrow R}\to\sqrt{\Xi\over|2\omega_c|}
\begin{cases}
{\overleftarrow B} e^{-i\omega_h r^*}&r\to r_h\\
e^{-i\omega_c r^*}+{\overleftarrow A}e^{i\omega_c r^*}&r\to r_c,
\end{cases}
\end{equation}
where the normalisation is discussed in appendix \ref{app:Normalization}.

Flux conservation follows from the Wronskian relation for any two solutions $R_1$ and $R_2$,
\begin{equation} \label{Wronskian}
\partial_{r^*}\left( R_1\partial_{r^*}R_2-R_2\partial_{r^*}R_1\right)=0.
\end{equation}
It leads to
\begin{align}
\omega_h(1-|\overrightarrow A_{\omega lm_{z}}|^2)&=\omega_c|\overrightarrow B_{\omega lm_{z}}|^2\\
\omega_c(1-|\overleftarrow A_{\omega lm_{z}}|^2)&=\omega_h|\overleftarrow B_{\omega lm_{z}}|^2\\
\omega_h\overrightarrow B_{\omega lm_{z}}\overrightarrow A^*_{\omega lm_{z}}&=-
\omega_c\overrightarrow B_{\omega lm_{z}}^*\overrightarrow A_{\omega lm_{z}}\\
\omega_h\overleftarrow B_{\omega lm_{z}}&=\omega_c\overrightarrow B_{\omega lm_{z}}.\label{flux}
\end{align}
From these we also conclude that $|\overrightarrow A_{\omega lm_{z}}|=|\overleftarrow A_{\omega lm_{z}}|$, 
and we define the absolute value of the reflection coefficient as 
$A_{\omega l m_{z}} \equiv |\overrightarrow A_{\omega lm_{z}}|=|\overleftarrow A_{\omega lm_{z}}|$.

\begin{figure} 
\centering
\includegraphics[width=0.4\linewidth]{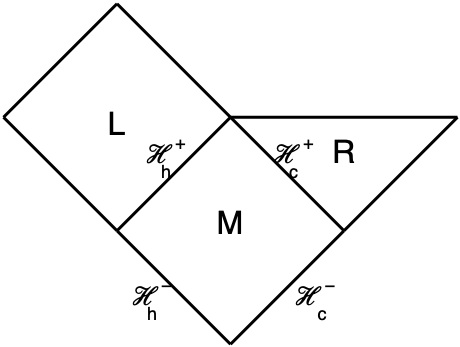}
\caption
{Part of the Penrose diagram of the black hole spacetime with the supported regions for
each of the mode function. For example, $f^M$ is based on support on region $M$.
} 
\label{fig:modes}
\end{figure}

In true Unruh style \cite{PhysRevD.14.870}, a superscript $L$, $M$ or $R$ denotes modes with support
in different parts of the extended Penrose diagram:
\begin{align}
L:\quad&r<r_h,\quad U_h=e^{-\kappa_hu}\\
M:\quad&r_h<r<r_c,\quad U_h=-e^{-\kappa_hu},\quad V_c=-e^{-\kappa_cv}\\
R:\quad&r_c<r,\quad V_c=e^{-\kappa_cv}
\end{align}
where $u= t-r^{\ast}$ and $v=t+r^{\ast}$ (see figure \ref{fig:modes}).
The $\kappa_{h,c}$ are the surface gravities at the respective horizons which are 
related to temperature in the usual manner, namely $T_{h,c} = \kappa_{h,c}/2\pi$.
(Note that we have taken positive surface gravities.)
The explicit expression for the temperatures are,
\begin{equation}
T_{h,c} \equiv \beta_{h,c}^{-1} =\left| \frac{\partial_r\Delta_r|_{r=r_{h,c}}}{4\pi(r_{h,c}^2+a^2)}\right|.
\end{equation}

The Kruskal coordinate on the event horizon is $U_h$ and the Kruskal coordinate on the
cosmological horizon is $V_c$. Let 
$f=R(r)e^{-i\omega t}S_{\omega lm_{z}}(\theta) e^{im_{z} \varphi}/\sqrt{r^2+a^2}$,
then outgoing modes which are positive frequency in $U$
on the past event horizon are denoted by $\overrightarrow f^U$,
\begin{equation}
\overrightarrow f^U=\left|2\sinh \pi\omega_h/\kappa_h\right|^{-1/2}
\left(e^{\pi\omega_h/2\kappa_h}\overrightarrow f^M+
e^{-\pi\omega_h/2\kappa_h}\overrightarrow f^L\right),
\end{equation}
where $\overrightarrow f^M$ and $\overrightarrow f^L$ are outgoing plane-wave basis 
of positive frequency defined in $U_{h}<0$ and $U_{h} > 0$, respectively.
Ingoing modes which are positive frequency in $V$ on the past cosmological horizon 
are denoted by $\overleftarrow f^V$,
\begin{equation}
\overleftarrow f^V=\left|2\sinh \pi\omega_c/\kappa_c\right|^{-1/2}
\left(e^{\pi\omega_c/2\kappa_c}\overleftarrow f^M+
e^{-\pi\omega_c/2\kappa_c}\overleftarrow f^R\right)
\end{equation}

We can check that the $\overrightarrow f^U$ modes are positive frequency
by taking a point on the horizon at $r=r_h$,
\begin{equation}
\overrightarrow f^U\to\sqrt{\Xi\over|4\omega_h\sinh \pi\omega_h/\kappa_h|}
\left(e^{-i\omega_h (t-r^*)+\pi\omega_h/2\kappa_h}\Theta(-U_h)+
e^{-i\omega_h (t-r^*)-\pi\omega_h/2\kappa_h}\Theta(U_h)\right)
\end{equation}
where $\Theta$ is the Heaviside function. In terms of the Kruskal coordinate, we have
\begin{equation}
\overrightarrow f^U\to\sqrt{\Xi\over|4\omega_h\sinh \pi\omega_h/\kappa_h|}
\exp\left({i\omega_h\over\kappa_h}\log(-iU_h)\right)
\end{equation}
Since the branch cut for the logarithm lies in the upper half plane,
the modes are regular in the entire lower half of the complex $U_h$ plane.
The Fourier transform in $U_h$ therefore vanishes when the Fourier frequency
is negative.

\section{Fluxes}
\label{sec:flux}

In order to employ quantum field theory on a curved spacetime background, we assume that the 
backreaction effect is small, and the expectation value of the quantum energy flux 
$\langle \hat T_{rt}\rangle$ reacts back on the spacetime geometry through the Einstein field equation
\begin{equation}
G_{rt}=8\pi \langle \hat T_{rt}\rangle.\label{qftics}
\end{equation}
Following previous work 
\cite{Hawking:1974rv,Hawking:1974sw,Page:1976df,Page:1976ki,PhysRevD.21.2185}, 
we also adopt an adiabatic approximation that assumes the
spacetime metric approximates the original black hole form but with time-dependent mass
and angular momentum. Note that, in the rotating case, the flux can introduce additional 
distortions dependent on the azimuthal angle, but we know that such distortions
decay away rapidly when not driven, and therefore we do not expect to see secular growth.
We remove these small distortions by averaging over the azimuthal angle.

For a non-minimally coupled scalar field, the stress energy tensor
\begin{equation}
T_{rt}=(1-8\pi\xi\phi^2)^{-1}\left\{\partial_r\phi\partial_t\phi-\xi\partial_t\partial_r\phi^2\right\}.
\end{equation}
Since the backreaction is small, and $\langle\hat \phi\rangle=0$, only the quadratic order terms need be
retained. Following Candelas \cite{PhysRevD.21.2185}, we introduce a stress tensor bilinear
\begin{equation}
T_{rt}(f_1,f_2)=\frac12\mathrm{Re}\left[\partial_t f_1^*\partial_r f_2+
\partial_r f_1^*\partial_t f_2-4\xi\partial_r\partial_t (f_1^*f_2)\right]
\end{equation}
and then the expectation value of the quantum flux operator in the vacuum state is
\begin{equation}
\langle T_{rt}\rangle={1\over 4\pi}\sum_{lm_{z}}\int_{-\infty}^\infty{d\omega\over 2\pi}\left\{
T_{rt}(\overrightarrow f^U,\overrightarrow f^U)+T_{rt}(\overleftarrow f^V,\overleftarrow f^V)\right\}.
\end{equation}
In the rotating case, this will depend on the azimuthal angle $\theta$. We include
the additional average over $\theta$ in the definition of $\langle\langle\rangle\rangle$ 
to remove this dependence.

Consider the flux as $r\to r_c$, inserting the explicit expressions for the modes from Eq.\ (\ref{modes}),
\begin{equation}
\langle\langle T_{rt}\rangle\rangle=-{\Xi\over 4\pi \Delta_r}\sum_{lm_{z}}\int_{-\infty}^\infty {d\omega\over 2\pi}
{\omega\over 2}
\left\{
{\omega_c\over\omega_h} |\overrightarrow B_{\omega lm_{z}}|^2
{e^{\beta_h\omega_h}\over e^{\beta_h\omega_h}-1}
-(1-|A_{\omega lm_{z}}|^2) {e^{\beta_c\omega_c}\over e^{\beta_c\omega_c}-1}
\right\}.
\end{equation}
After using the flux rules Eqs.\ (\ref{flux}) and the symmetry $A_{-\omega l -m}=A_{\omega l m}$,
\begin{equation} \label{Trt}
\langle\langle T_{rt}\rangle\rangle=-{\Xi\over 4\pi \Delta_r}\sum_{lm_{z}}\int_0^\infty {d\omega\over 2\pi}\omega
\left(1-|A_{\omega lm_{z}}|^2\right)\frac12\left\{{\rm coth}{\beta_h\omega_h\over 2}
-{\rm coth}{\beta_c\omega_c\over 2}\right\}.
\end{equation}
Due to the constancy of the Wronskian \eqref{Wronskian}, the expression under the integral is 
independent of the radius we choose to evaluate $T_{rt}$.
\red{The factor of $\Delta_r^{-1}$ comes from the derivative $dr^{\ast}/dr$ and the divergence of $1/\Delta_r$ at the cosmological horizon is relevant to the blue shift effect although the mass-loss and spin-loss rates are regular as is shown later.
Consequently, the only $r$ dependence appearing in \eqref{Trt} is contained in the prefactor $\Delta_r^{-1}$.}
For the rotation, consider $T_{r\varphi}$,
\begin{equation}
\langle\langle T_{r\varphi}\rangle\rangle={\Xi\over 4\pi \Delta_r}\sum_{lm_{z}}\int_0^\infty {d\omega\over 2\pi} m_z
\left(1-|A_{\omega lm_{z}}|^2\right)\frac12\left\{{\rm coth}{\beta_h\omega_h\over 2}
-{\rm coth}{\beta_c\omega_c\over 2}\right\}.
\end{equation}

The rate of mass loss can be found in an adiabatic approximation by replacing $m$ and $a$ 
by a functions $m(t)$ and $a(t)$, then the relevant Einstein tensor components averaged over 
angles becomes
\begin{align}
\langle\langle G_{rt}\rangle\rangle&={2\over \Delta_r\Xi}\left(\Xi\dot m-3{a\over\ell^2}m\dot a\right),\\
\langle\langle G_{r\varphi}\rangle\rangle&={2\over \Delta_r\Xi^2}\left(\left(3{a^2\over\ell^2}-1\right)m\dot a
-a\Xi \dot m\right),
\end{align}
Using $M=m/\Xi^2$ and $J=am/\Xi^2$ for the slowly evolving background spacetime,
\begin{align}
\langle\langle G_{rt}\rangle\rangle+{a\over\ell^2}\langle\langle G_{r\varphi}\rangle\rangle&=
\Xi{2\over \Delta_r}{dM\over dt},\\
\langle\langle G_{r\varphi}\rangle\rangle&=-\Xi{2\over \Delta_r}{dJ\over dt}.
\end{align}
Hence, relating these to the fluxes by Eq. (\ref{qftics}),
\begin{equation}
{dM\over dt}=-\sum_{lm_{z}}\int_0^\infty {d\omega\over 2\pi}\left(\omega-{am_z\over\ell^2}\right)
\left(1-|A_{\omega lm_{z}}|^2\right)\frac12\left\{{\rm coth}{\beta_h\omega_h\over 2}
-{\rm coth}{\beta_c\omega_c\over 2}\right\}.\label{massloss}
\end{equation}
and for the angular momentum,
\begin{equation}
{dJ\over dt}=-\sum_{lm_{z}}\int_0^\infty {d\omega\over 2\pi}m_z
\left(1-|A_{\omega lm_{z}}|^2\right)\frac12\left\{{\rm coth}{\beta_h\omega_h\over 2}
-{\rm coth}{\beta_c\omega_c\over 2}\right\}.\label{amloss}
\end{equation}
Note that due to the definition of $\Omega$ in \eqref{ang_vel}, the factor $\omega-am_z/\ell^2$ in \eqref{massloss} is equivalent to $\omega-m_z\Omega(\ell)$, i.e. the frequency in the rotating frame at $r=\ell$.
Indeed, in black hole thermodynamics in de Sitter space, the angular velocities appearing in the first laws for the two horizons are $\Omega_{h,c}=\Omega(r_{h,c})-\Omega(\ell)$ \cite{Dolan:2013ft}.
Thus, one would expect the relevant frequency for the change in thermodynamic mass $M$ to be the one that appears in \eqref{massloss}.

We have been assuming that the energy fluxes affect the mass and spin of the black hole,
but there has been some discussion elsewhere about whether quantum processes could cause
the decay of the vacuum energy in de Sitter space 
\cite{Mottola:1984ar,Abramo:1997hu,Anderson:2013ila,Markkanen:2016jhg}. 
In order to examine this possibility, we let $\ell\to\ell(t)$
and examine the angle averaged $G_{rt}$ Einstein tensor component,
\begin{equation}
\langle G_{rt}\rangle =2{\dot m\over\Delta_r}-{6ma\over\ell^2\Xi}{\dot a\over\Delta_r}
+{6ma^2\over\ell^3\Xi}{\dot \ell\over\Delta_r}-{(a^2+r^2)r\over \ell^3}{\dot \ell\over\Delta_r}.
\end{equation}
The first three terms match the radial dependence in the fluxes, but the final term
does not. We conclude that decay of vacuum energy is not consistent with the backreaction
calculation in the vacuum state that we are using.

\section{Evaporation process of a spinning BH with a cosmological constant}\label{evaporation}

In this section, we describe how we compute numerically the time-development of a 
spinning black hole with a cosmological constant. 

\subsection{Methodology}

Following the earlier work of Page \cite{Page:1976ki}, we
assume the angular momentum $J(t)$ is monotonically decreasing, and choose the spin parameter 
$a$ as the independent variable characterizing the evolution of a rotating black hole, i.e.
\begin{equation}
M (t) \ \text{and} \ J(t) \to M(a) \ \text{and} \ t(a).
\end{equation}
A large dynamical range can be investigated using logarithmic variables,
\begin{align}
y &\equiv -\ln{(a/a_{0})},\\
z &\equiv -\ln{(M/M_0)},
\end{align}
where $a_{0}$ and $M_0$ are the initial spin parameter and mass. 
The function $z(y)$ evolves as
\begin{equation}
\frac{dz}{dy} = \frac{f}{g-f}.
\label{z_evo}
\end{equation}
where
\begin{align}
f&= -M^{3} \frac{d \ln{(M/M_{0})}}{dt},\\
g&= -M^{3} \frac{d \ln{(J/J_{0})}}{dt},
\end{align}
The mass and angular momentum loss rates are evaluated numerically using (\ref{massloss})
and (\ref{amloss}).

Finally, the scale-invariant time parameter, $\tau$, is defined as
\begin{equation}
\tau \equiv \frac{t}{M_{0}^{3}},
\end{equation}
and the evolution of $\tau (y)$ is governed by
\begin{equation}
\frac{d\tau}{dy} = \frac{e^{-3z}}{g-f}.
\label{tau_evo}
\end{equation}
We solve the two ordinary differential equations (ODEs), (\ref{z_evo}) and (\ref{tau_evo}), with the 
fourth-order Runge-Kutta method while the values of $f$ and $g$ are obtained as explained below.

\subsection{Teukolsky equation and the Heun function}

For the analysis of the linear perturbations of gravitational, electromagnetic, 
neutrino, and scalar fields, it is important to estimate the greybody factor. Let us 
consider a mode function ${}_{s}\psi_{l m_{z}} (t, r, \theta, \varphi)$, 
where $s$ is the spin of the field, which can be expanded as
\begin{equation}
{}_{s} \psi_{l m_{z}} (t, r, \theta, \varphi) = \int \frac{d \omega}{2 \pi} e^{-i \omega t} e^{i m_{z} \varphi} {}_{s}R^{\ct}_{\omega l m_{z}} (r) {}_{s}S_{\omega l m_{z}} (\theta).
\end{equation}
Note that the radial function $R_{\omega l m_{z}}$ introduced in (\ref{expand}) is related to ${}_s R_{\omega l m_z}^{\ct}$ as $R_{\omega l m_{z}} = (r^{2}+a^{2})^{1/2} {}_{0} R^{\ct}_{\omega l m_{z}}$.
It was shown that the functions ${}_{s}R^{\ct}_{\omega l m_{z}} (r)$ and ${}_{s}S_{\omega l m_{z}} (\theta)$ are solutions of the following Teukolsky equations by Suzuki, Takasugi, and Umetsu in Ref. \cite{Suzuki:1998vy}:
\begin{align}
\left[ \frac{d}{dq} \Delta_q \frac{d}{dq} - \frac{1}{\Delta_q} \left(  \tilde{V} + \frac{s}{2} \Delta_q' \right)^2 + 2s \tilde{V}' - {}_{s}X_{\omega lm_{z}} \right] {}_{s} S_{\omega l m_{z}} &= 0, \label{angular_T}\\
\left[ \Delta_r^{-s} \frac{d}{dr} \Delta_r^{s+1} \frac{d}{dr} + \frac{1}{\Delta_r} \left( \tilde{W}^2 -is\tilde{W} \Delta_r' \right) + 2is \tilde{W}' - {}_{s}Y_{\omega lm_{z}} \right]{}_{s} R^{\ct}_{\omega l m_{z}} &= 0, \label{radial_T}
\end{align}
where
\begin{align}
\tilde{V}(q) &\equiv a\omega (1-q^2) - \Xi m,\\
\tilde{W}(r) &\equiv \omega (r^2+a^2) -\Xi am,\\
{}_{s}X_{\omega lm_{z}}(q) &\equiv 2 (2s^2 + 1) \alpha^2 q^2 - {}_s \lambda_{\omega lm_{z}},\\
{}_{s}Y_{\omega lm_{z}}(r) &\equiv 2 \ell^{-2} (s+1) (2s+1) r^2 + {}_s \lambda_{\omega lm_{z}} -s (1-\alpha^2),\\
\Delta_q &\equiv (1-q^2) (1+\alpha^2 q^2),\\
q &\equiv \cos \theta,\\
\alpha &\equiv a/\ell,
\end{align}
and ${}_s \lambda_{\omega lm_{z}}$ is the angular eigenvalue. The prime in $\Delta_q' (q)$ and $\Delta_r' (r)$ denotes the derivative with respect to $q$ and $r$, respectively. Note that the Teukolsky equations with $s=0$ describe perturbations of a massless conformally coupled scalar field. In the following,  for scalar perturbations, we will deal with the conformally coupled scalar field ($\xi =1/6$) only. The angular eigenvalue ${}_s \lambda_{\omega lm_{z}}$ is determined by a boundary condition requiring the regularity for ${}_{s}S_{\omega l m_{z}}$ at $q=\pm 1$. As shown in \cite{Suzuki:1998vy}, this boundary value problem can be solved by the Heun function. Let us implement the following transformations in (\ref{angular_T}):
\begin{align}
z &\equiv \frac{(1-i/\alpha) (1+q)}{2 (q-i/\alpha)},\\
{}_{s}\tilde{S}_{\omega l m_{z}} (z) &\equiv z^{-A_1} (z-1)^{-A_2} (z-z_a)^{-A_3} (z-z_{a\infty})^{-1} {}_{s} S_{\omega l m_{z}} (q),
\end{align}
where 
\begin{align}
z_{a} &\equiv - \frac{(1-i/\alpha)^2}{4 i/\alpha},\\
z_{a\infty} &\equiv \frac{1-i / \alpha}{2},
\end{align}
and
\begin{align}
A_1 \equiv \frac{m-s}{2}, \ \ \ A_{2} \equiv - \frac{m+s}{2}, \ \ \ A_3 \equiv \frac{i}{2} \left( \frac{c}{\alpha} - m \alpha -is \right)
\end{align}
with $c\equiv a \omega$. In the following, we often omit the subscripts $(s,\omega,l,m_{z})$. The function $\tilde{S} (z)$ satisfies the following ODE
\begin{equation}
\tilde{S}''+ \left( \frac{2A_1+1}{z} + \frac{2A_2+1}{z-1} + \frac{2A_3 + 1}{z-z_{a}} \right) \tilde{S}'
+ \frac{F_a^{(+)} F_a^{(-)} z + G_a}{z (z-1) (z-z_{a})} \tilde{S} = 0,
\label{Heun_angular}
\end{equation}
where a prime denotes the derivative with respect to $z$ and
\begin{align}
F_a^{(+)} &\equiv 1, \  F_a^{(-)} \equiv 1-s-im \alpha + i \frac{c}{\alpha},\\
G_a &\equiv - \left[ \frac{i \lambda}{4 \alpha} + \frac{1}{2} +A_1 + \left( m+\frac{1}{2} \right) (A_3-A_3^{\ast}) \right].
\end{align}
Eq.\ (\ref{Heun_angular}) is nothing but Heun's differential equation. One can obtain the analytic solutions at $z=0$ ($q=-1$) and $z=1$ ($q=+1$) with the general Heun function denoted by ``HeunG''. Indeed, the radial part (\ref{radial_T}) can be also transformed into Heun's differential equation. Let us implement the M$\ddot{\rm o}$bius transformation
\begin{equation}
z = \frac{(r_c-r_h') (r-r_h)}{(r_c-r_h) (r-r_h')},
\end{equation}
and define a new variable
\begin{equation}
\tilde{R}^{\ct} (z) \equiv z^{-B_1} (z-1)^{-B_2} (z-z_r)^{-B_3} (z-z_{r\infty})^{-2s-1} R^{\ct}(r),
\end{equation}
with
\begin{align}
&B_1 \equiv \frac{i \tilde{W} (r_h)}{\Delta_r'(r_h)}, \ \ \ B_2 \equiv \frac{i \tilde{W} (r_c)}{\Delta_r'(r_c)}, \ \ \ 
B_3 \equiv \frac{i \tilde{W} (r_c')}{\Delta_r'(r_c')},\\
&z_r \equiv \frac{(r_c-r_h') (r_c' - r_h)}{(r_c - r_h) (r_c' - r_h')}, \ \ \ z_{r\infty} \equiv \frac{r_c - r_h'}{r_c-r_h}.
\end{align}
Here $r_{c}'$, $r_{h}'$, $r_{h}$, and $r_{c}$ are four roots of $\Delta_r (r)= 0$ and they satisfy $r_c' < r_h' < r_h < r_c$. Note that the outer horizon at $r=r_{h}$ and cosmological horizon at $r=r_{c}$ correspond to $z=0$ and $z=1$, respectively. Then we obtain Heun's differential equation
\begin{align}
\begin{split}
\tilde{R}^{\ct} {}'' + \left( \frac{2B_1+s+1}{z} + \frac{2B_2+s+1}{z-1} + \frac{2B_3+s + 1}{z-z_r} \right) \tilde{R}^{\ct}{}'&\\
+ \frac{F_r^{(+)} F_r^{(-)} z + G_r}{z (z-1) (z-z_r)} \tilde{R}^{\ct} = 0&,
\end{split}\label{Heun_radial}
\end{align}
where
\begin{align}
F_r^{(+)} &\equiv 2s+1, \ F_r^{(-)}\equiv s+1- \frac{2i \tilde{W}(r_h')}{\Delta_r'(r_h')},\\
\begin{split}
G_r &\equiv \frac{(1+s) (1+2s) r_c'}{r_h' - r_c'} + \frac{\ell^2[\lambda-2s (1-\alpha^2)]+(1+s) (1+2s) r_h (r_c+r_h)}{ (r_h'-r_c') (r_h-r_c)}\\
&-\frac{2i\ell^2 \Xi (1+2s) (r_h r_h' \omega +a^2 \omega/\Xi -am)}{ (r_h'-r_c') (r_h'-r_h) (r_h-r_c)}.
\end{split}
\end{align}

Heun's differential equation in a general form is
\begin{equation}
y'' (z) + \left( \frac{\gamma}{z} + \frac{\delta}{z-1} + \frac{\epsilon}{z-k} \right) y'(z) + \frac{\zeta \beta z -w}{z (z-1) (z-k)} y(z) = 0,
\end{equation}
with $\gamma + \delta + \epsilon = \zeta + \beta + 1$. Heun's equation has four regular singular points at $z=0$, $z=1$, $z=k$, and $z = \infty$. One can construct local solutions for each singular point, and those solutions are convergent inside a circle. The radius of the convergence is determined by the distance from the neighbour singular point. In our analysis, we need two local solutions at $z=0$ and $z=1$. The general solution near $z=0$ is given by the linear combination of \cite{Hatsuda:2020sbn}
\begin{align}
y_{01} (z) &= \text{HeunG} (k,w;\zeta , \beta , \gamma , \delta; z),\\
y_{02} (z) &= z^{1-\gamma} \text{HeunG} (k, (k \delta+\epsilon) (1-\gamma) +w; \zeta+1-\gamma , \beta + 1-\gamma, 2-\gamma , \delta; z),
\end{align}
and the general solution near $z = 1$ is represented by the linear combination of
\begin{align}
y_{11} (z) &= \text{HeunG} (1-k,\zeta \beta - w;\zeta , \beta , \delta , \gamma; 1-z),\\
\begin{split}
y_{12} (z) &= (1-z)^{1-\delta} \\
&\times \text{HeunG} (1-k, p; \zeta+1-\delta , \beta + 1-\delta, 2-\delta , \gamma; 1-z),
\end{split}
\end{align}
where $p\equiv ((1-k) \gamma + \epsilon) (1-\delta) + \zeta \beta -w$.
Let the function $y_{Aij} (z)$ be defined as $y_{ij} (z)$ with the values
\begin{align}
\begin{split}
&\gamma \to 2A_1+1, \ \delta \to 2A_2+1, \ \epsilon \to 2A_3+1,\\
&\zeta \to F_a^{(+)}, \ \beta \to F_a^{(-)}, \ w \to G_a,
\end{split}
\end{align}
and $y_{Rij} (z)$ is defined by $y_{ij} (z)$ with
\begin{align}
\begin{split}
&\gamma \to 2B_1+1, \ \delta \to 2B_2+1, \ \epsilon \to 2B_3+1,\\
&\zeta \to F_r^{(+)}, \ \beta \to F_r^{(-)}, \ w \to G_r.
\end{split}
\end{align}
Then the two solutions of the angular equation near $z=0$ ($q=-1$) is given by
\begin{align}
S_{01} (q) &\equiv z^{A_{1}} (z-1)^{A_{2}} (z-z_{a})^{A_{3}} (z-z_{\infty}) y_{A01}(z) \propto (1+u)^{(m_{z}-s)/2},\\
S_{02} (q) &\equiv z^{A_{1}} (z-1)^{A_{2}} (z-z_{a})^{A_{3}} (z-z_{\infty}) y_{A02}(z) \propto (1+u)^{-(m_{z}-s)/2},
\end{align}
and near $z=1$ ($q=+1$), we have
\begin{align}
S_{11} (q) &\equiv z^{A_{1}} (z-1)^{A_{2}} (z-z_{a})^{A_{3}} (z-z_{\infty}) y_{A11}(z) \propto (1-u)^{-(m_{z}+s)/2},\\
S_{12} (q) &\equiv z^{A_{1}} (z-1)^{A_{2}} (z-z_{a})^{A_{3}} (z-z_{\infty}) y_{A12}(z) \propto (1-u)^{(m_{z}+s)/2}.
\end{align}
In order for the solution to be regular at $q=\pm1$, one has to properly choose the boundary solutions at $q=\pm1$. As an example, let us consider the case of $(s,m_{z}) = (1,0)$. In this case, the solution at $q=-1$ and $q=+1$ should be $\sim S_{02} (q)$ and $\sim S_{12} (q)$, respectively, in order for the solution to be regular. Such a situation is realized when those two solutions are linearly dependent, which is equivalent to
\begin{equation}
W[y_{A02}, y_{A12}] = 0,
\end{equation}
and this condition determines the angular eigenvalue $\lambda_{slm}$. We search the exact value of $\lambda_{slm_{z}}$ by using the function \texttt{FindRoot} in \textit{Mathematica}.

\subsection{Greybody factor and superradiance}
We calculate the greybody factor by using the analytic solutions of the Teukolsky equation, represented by the general Heun function. 
Using the analytic solutions of Heun's equation at $z=0$ and $z=1$, the two independent radial solutions near the outer horizon $r=r_h$ ($z=0$)
\begin{align}
R_{+}^{(h)} &\equiv z^{B_{1}} (z-1)^{B_{2}} (z-z_{r})^{B_{3}} (z-z_{\infty})^{2s+1} y_{R01}(z),\\
R_{-}^{(h)} &\equiv z^{B_{1}} (z-1)^{B_{2}} (z-z_{r})^{B_{3}} (z-z_{\infty})^{2s+1} y_{R02}(z),
\end{align}
and near the cosmological horizon $r=r_{c}$ ($z=1$), we have
\begin{align}
R_{+}^{(c)} &\equiv z^{B_{1}} (z-1)^{B_{2}} (z-z_{r})^{B_{3}} (z-z_{\infty})^{2s+1} y_{R11}(z),\\
R_{-}^{(c)} &\equiv z^{B_{1}} (z-1)^{B_{2}} (z-z_{r})^{B_{3}} (z-z_{\infty})^{2s+1} y_{R12}(z).
\end{align}
In the near-horizon limit, those solutions reduce to purely ingoing or outgoing modes
\begin{align}
R_{\pm}^{(h)} \simeq C_{\pm}^{(h)} (r-r_h)^{-\frac{s}{2} \pm \theta_{h}}, \ \theta_{h} \equiv i\frac{\omega_{h}}{2\kappa_{h}}+\frac{s}{2},\\
R_{\pm}^{(c)} \simeq C_{\pm}^{(c)} (r-r_{c})^{-\frac{s}{2} \pm \theta_{c}}, \ \theta_{c} \equiv i\frac{\omega_{c}}{2\kappa_{c}}+\frac{s}{2},
\end{align}
where $\kappa_{h/c} \equiv 2\pi/\beta_{h/c}$ and
\begin{align}
C_{\pm}^{(h)} &\equiv(-1)^{B_{2}} \left( \frac{r_c-r_h'}{(r_c- r_h) (r_h -r_h')} \right)^{-\frac{s}{2} \pm \theta_{h}} \left( \frac{(r_c - r_h') (r_h - r_c')}{(r_c -r_h) (r_c' - r_h')} \right)^{B_{3}} \left(\frac{r_h' - r_c}{r_c -r_h} \right)^{2s+1},\\
C_{\pm}^{(c)} & \equiv \left( \frac{r_h- r_h'}{(r_c - r_h) (r_c -r_h')} \right)^{-\frac{s}{2} \pm \theta_{c}} \left( \frac{(r_h -r_h') (r_c - r_c')}{(r_c -r_h) (r_c' -r_h')} \right)^{B_{3}} \left( \frac{r_h' -r_h}{r_c -r_h} \right)^{2s+1}.
\end{align}

\begin{figure} 
\centering
\includegraphics[width=1\linewidth]{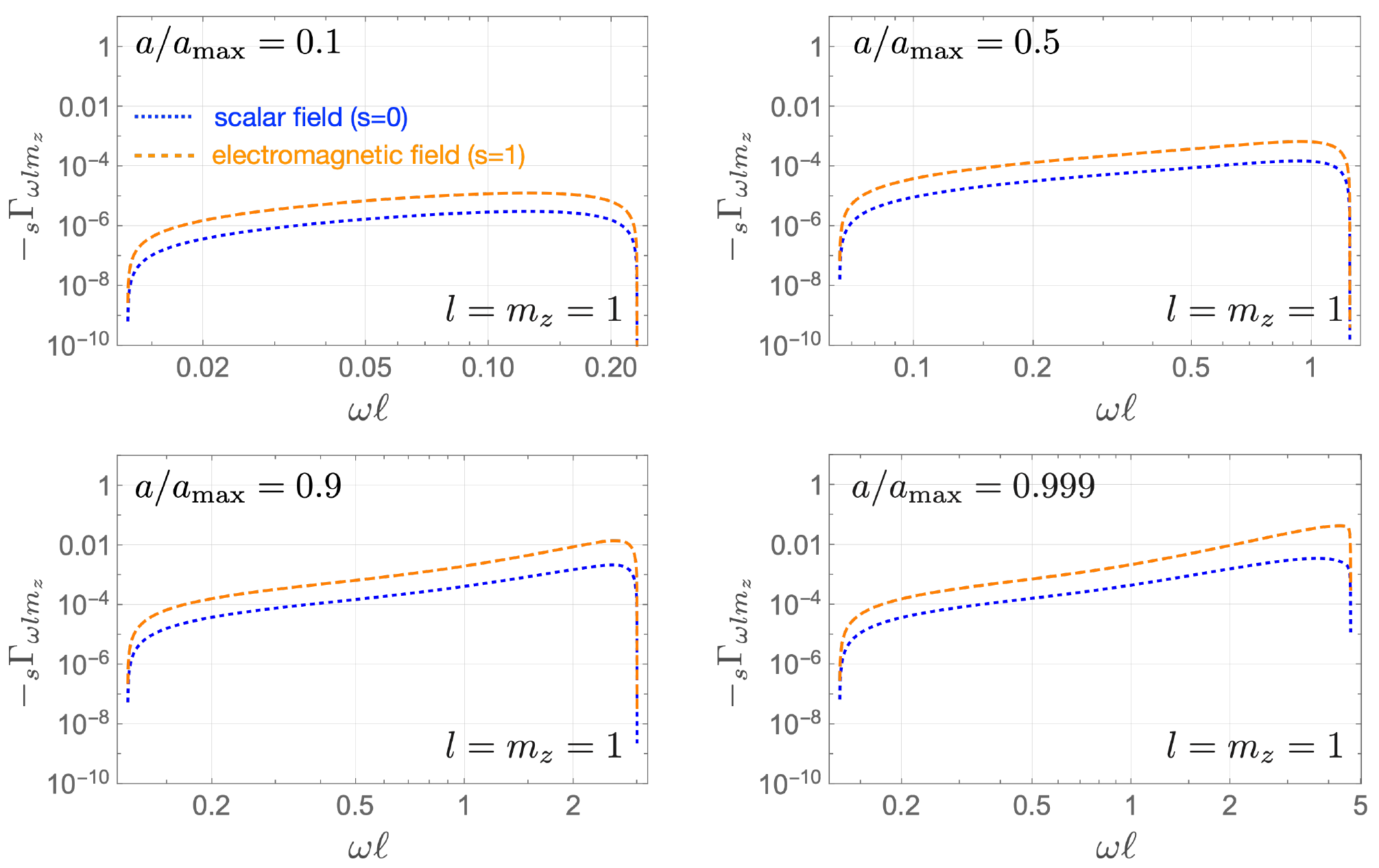}\hspace{1cm}
\includegraphics[width=1\linewidth]{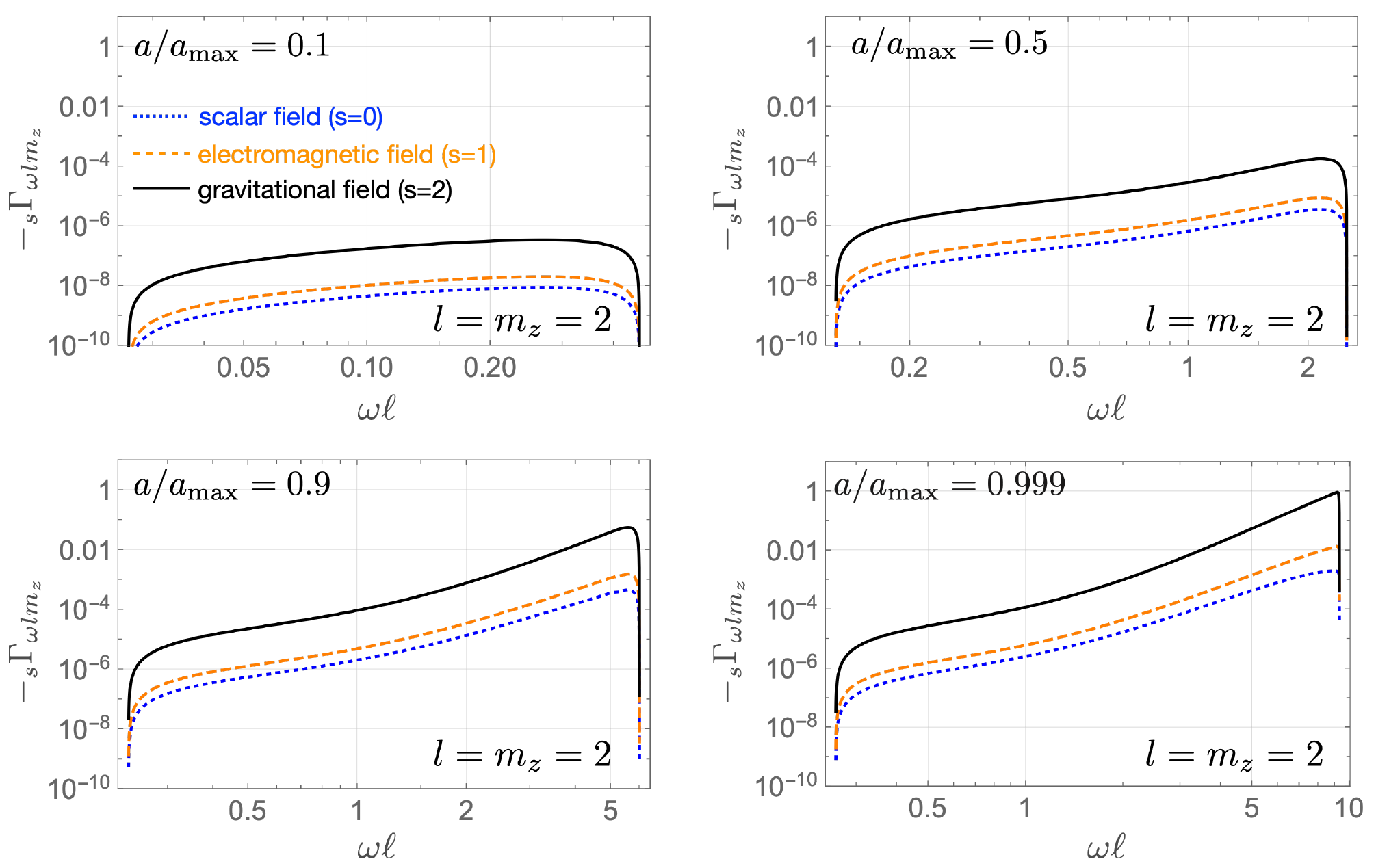}\hspace{1cm}
\caption{Plot of $-{}_{s}\Gamma_{\omega lm_{z}}$ in the frequency range showing superradiance. 
We set $m/\ell=0.1$ and $a_{\rm max}$ is the spin parameter at the extremal situation with $m/\ell=0.1$. 
} \label{fig:SR}
\end{figure}
Let us consider the solution for which the mode function is purely outgoing at $r=r_{c}$
\begin{align}
{}_{s}R^{\ct}_{\omega l m_{z}} = 
\begin{cases}
{}_{s}B_{\omega l m_{z}} (r-r_{c})^{i \frac{\omega_{c}}{2 \kappa_{c}}} & r \to r_{c},\\
\displaystyle
(r-r_h)^{i \frac{\omega_{h}}{2 \kappa_{h}}} + {}_{s}A_{\omega l m_{z}} (r-r_h)^{-i \frac{\omega_{h}}{2 \kappa_{h}} -s}& r \to r_{+},
\end{cases}
\end{align}
and the coefficients are given by
\begin{align}
\frac{1}{{}_{s}B_{\omega l m_{z}}} = \frac{C_{+}^{(h)}}{C_{+}^{(c)}} \frac{W[y_{11},y_{02}]}{W[y_{01},y_{02}]},\label{ref_trans_1}\\
\frac{{}_{s}A_{\omega l m_{z}}}{{}_{s}B_{\omega l m_{z}}} = \frac{C_{-}^{(h)}}{C_{+}^{(c)}} \frac{W[y_{11},y_{01}]}{W[y_{02},y_{01}]}.
\label{ref_trans_2}
\end{align}
\begin{figure} 
\centering
\includegraphics[width=1\linewidth]{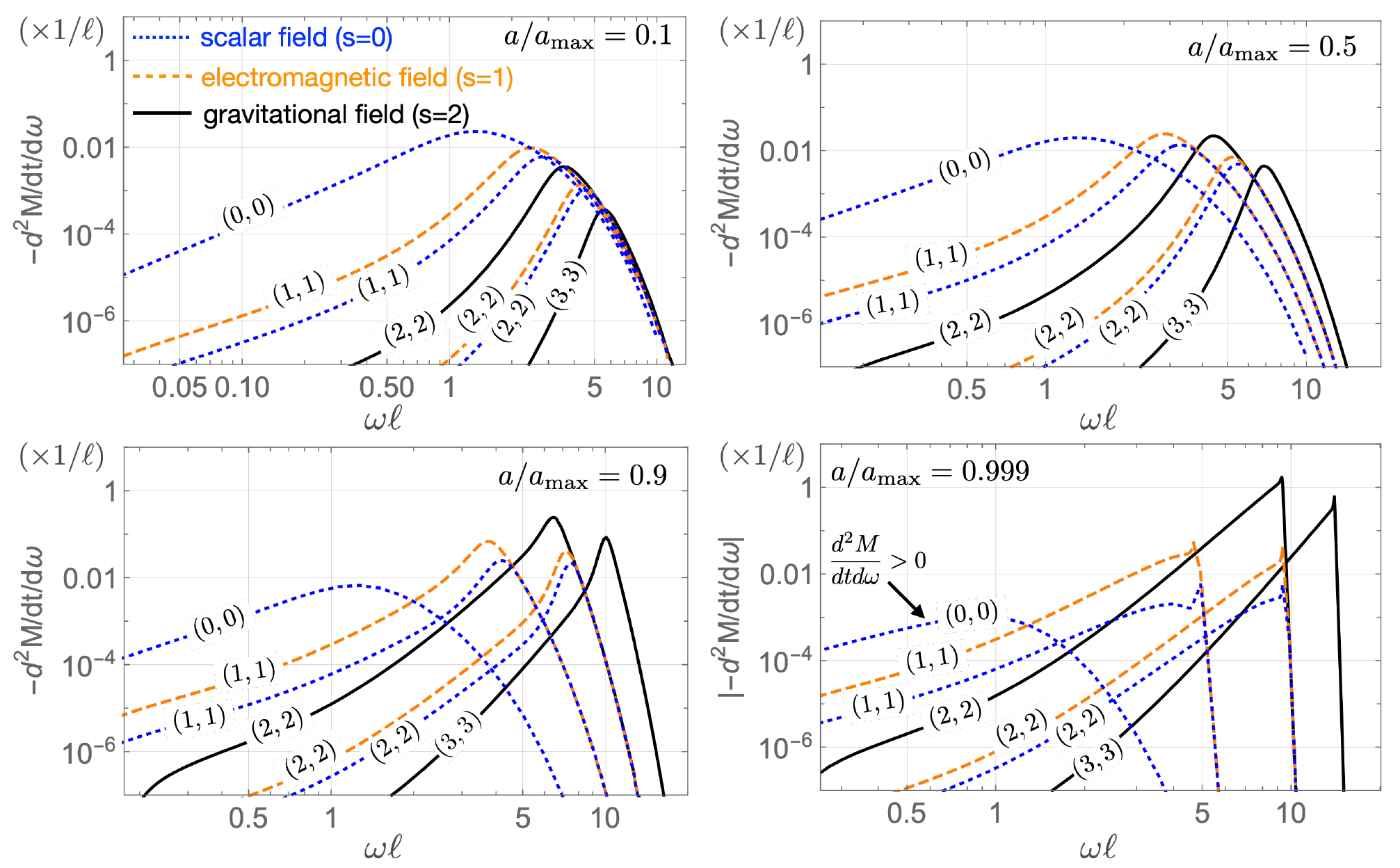}\hspace{1cm}
\caption{The spectra of energy fluxes carrying out the mass of KdS black hole. The spectra are computed with respect to various combinations of $(l,m_{z})$. The mass parameter is fixed with $m/\ell =0.1$. Note that the mass-change rate of the scalar component of $(0,0)$ mode for $a/a_{{\rm max}} = 0.999$ is positive although it is negligible.} \label{fig:spectrumM}
\end{figure}\nobreak
\begin{figure} 
\centering
\includegraphics[width=1\linewidth]{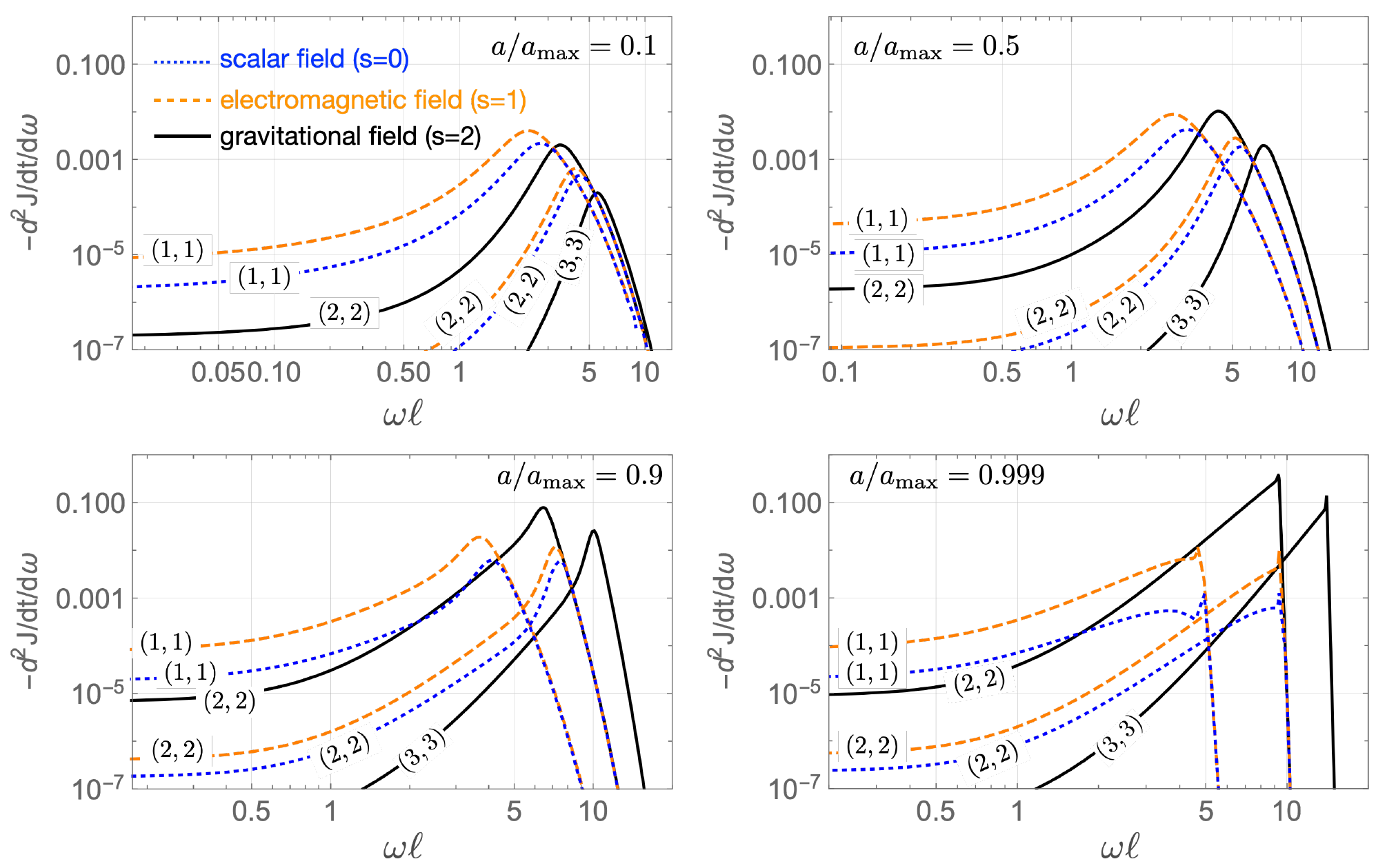}\hspace{1cm}
\caption{The spectra of energy fluxes carrying out the angular momentum of KdS black hole. The parameter set is same as figure \ref{fig:spectrumM}.} \label{fig:spectrumJ}
\end{figure}
Using the Wronskian for the solutions of Teukolsky equation
\begin{equation}
{}_{s}{\cal W} [u_{1}, u_{2}] \equiv -i \Delta_{r}^{s+1} (r) \left( u_{1} \frac{d u_{2}}{dr} - u_{2} \frac{d u_{1}}{d r} \right),
\label{Teukolsky_W}
\end{equation}
whose real part gives the energy flux, one can obtain the following relation \cite{Novaes:2018fry}
\begin{equation}
({}_{s}A_{\omega l m_{z}}) ({}_{-s}A_{\omega l m_{z}}^{\ast}) 
+ ({}_{s}\tau_{\omega lm_{z}}^{-1}) ({}_{s}B_{\omega l m_{z}}) 
({}_{-s}B_{\omega l m_{z}}^{\ast}) = 1,
\end{equation}
where the function ${}_{s}\tau_{\omega lm_{z}}$ is defined as
\begin{equation}
{}_{s} \tau_{\omega lm_{z}} \equiv \frac{\theta_{h} \Delta_{r}'(r_h)}{\theta_{c} \Delta_{r}'(r_{c})}.
\end{equation}
The Wronskian (\ref{Teukolsky_W}) is constant in $r$ when $u_{1}$ and $u_{2}$ are the solutions of the Teukolsky equation. Then the greybody factor, ${}_{s}\Gamma_{\omega l m_{z}}$, (transmissivity) is given by
\begin{equation}
{}_{s}\Gamma_{\omega l m_{z}} = 1- ({}_{s}A_{\omega l m_{z}}) 
({}_{-s}A_{\omega l m_{z}}^{\ast}) = ({}_{s} \tau_{\omega lm_{z}}^{-1} )
({}_{s}B_{\omega l m_{z}}) ({}_{-s}B_{\omega l m_{z}}^{\ast})\,.
\label{greybody}
\end{equation}
Now we can compute the greybody factor by using (\ref{ref_trans_1}), (\ref{ref_trans_2}), and (\ref{greybody}). Our high-accuracy computation shows the superradiance, which is quantified by the greybody factor. The greybody factors for (conformal-coupling) scalar, electromagnetic, and gravitational perturbations in the superradiance-frequency band, $m_{z} \Omega_{c} \leq \omega \leq m_{z} \Omega_{h}$, are shown in figure \ref{fig:SR}.

\subsection{Hawking spectrum}

We compute the spectrum of the energy fluxes for conformal scalar field, photon and graviton
emission. It was shown that the energy fluxes of the photon and graviton in the Kerr geometry 
are simply related to the Wronskians of the normalised modes to the appropriate Teukolsky equation
\cite{1974ApJ...193..443T,Chandrasekhar:1985kt}. \red{The analogous result for Kerr-de Sitter (KdS) can be found in 
Refs.\ \cite{Suzuki:1999pa,Novaes:2018fry}, where
it was shown that the constant Wronskian can be mapped into the conservation law of energy flux, from which one can implicitly derive the form of energy flux in KdS spacetime for a massless spin-$s$ field. They also confirmed that the greybody factor of KdS spacetime obtained in the Wronskian analysis is consistent with that of Kerr spacetime in the limit of $\ell \to \infty$. The fluxes are therefore determined by the greybody factors found above.}

\begin{figure} 
\centering
\includegraphics[width=1\linewidth]{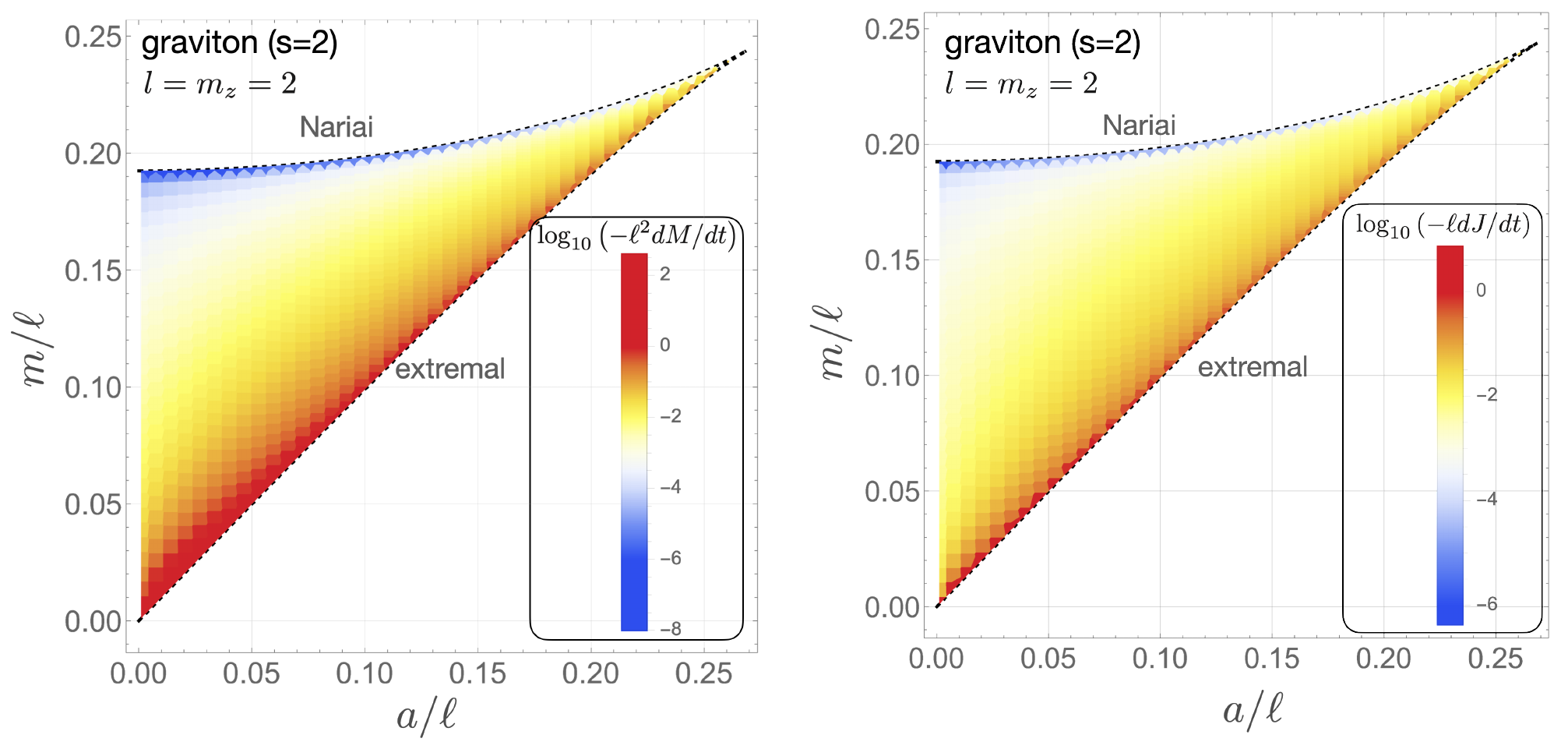}\hspace{1cm}
\caption{The mass-loss and spin-loss rates in the spin-mass parameter space. Here we computed the contribution of graviton-emission with $l=m_{z} =2$ only. The mass-loss and spin-loss rates at the Nariai limit vanish because of the zero net flux of Hawking radiation emitted from the black hole and cosmological horizons.
} \label{fig:flux_global}
\end{figure}

\begin{figure} [h]
\centering
\includegraphics[width=0.7\linewidth]{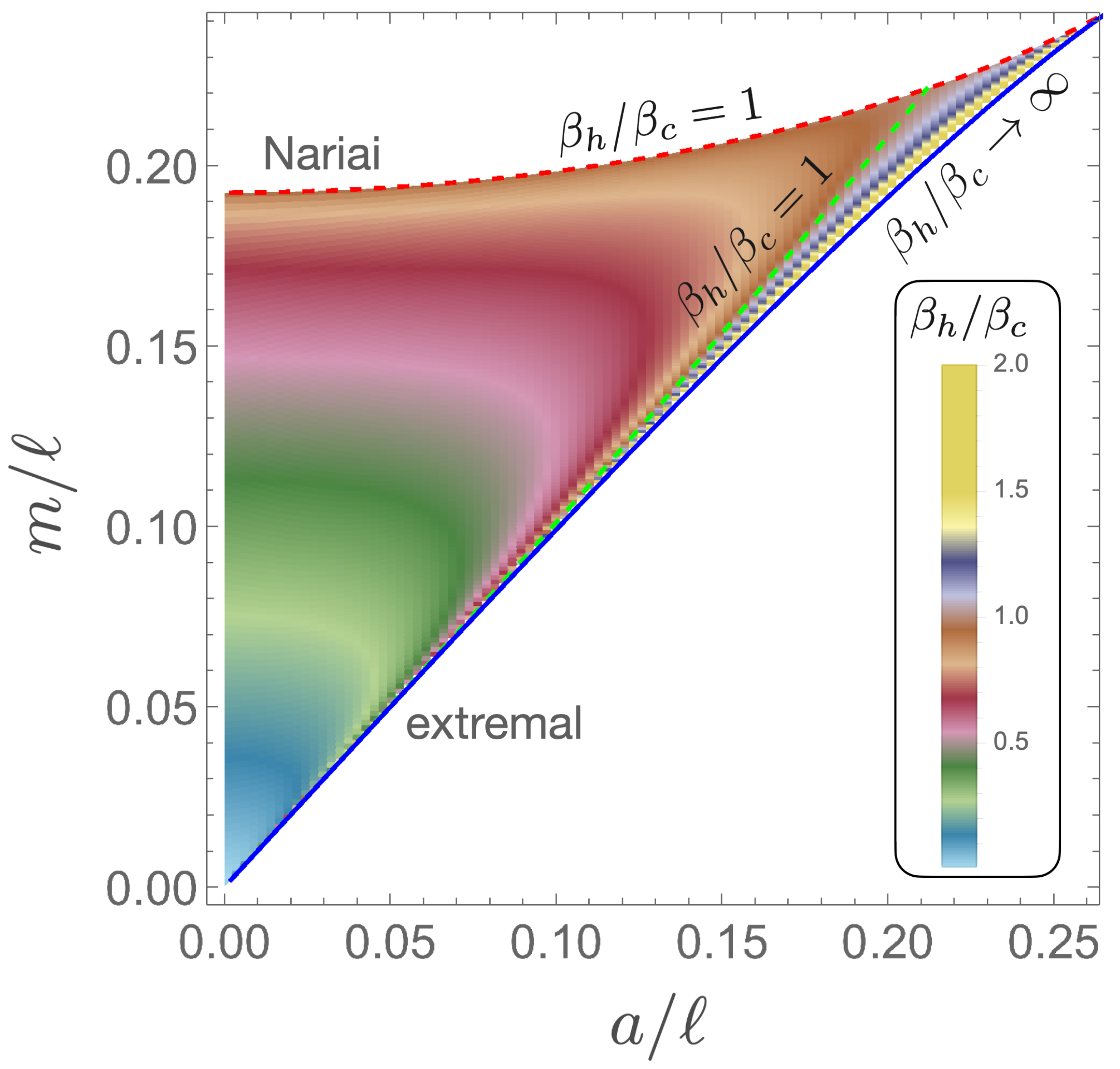}\hspace{1cm}
\caption{The ratio of the cosmological temperature to black hole temperature in the spin-mass parameter space. 
} \label{fig:ratio_temp}
\end{figure}

The evolution of the mass and angular momentum are given by,
\begin{equation}
\frac{d^{2}M}{dt d\omega}=-\sum_{slm_{z}}\frac{1}{2\pi} \left( \omega - a m_{z} /\ell^{2} \right) {}_{s}\Gamma_{\omega l m_{z}} \frac{1}{2} \left\{ \coth{\left(\frac{\beta_{h} \omega_{h}}{2} \right)} - \coth{\left(\frac{\beta_{c} \omega_{c}}{2} \right)} \right\},
\label{spectrumMM}
\end{equation}
and
\begin{equation}
\frac{d^{2}J}{dt d\omega}=-\sum_{slm_{z}}\frac{m_{z}}{2\pi} {}_{s}\Gamma_{\omega l m_{z}} \frac{1}{2} \left\{ \coth{\left(\frac{\beta_{h} \omega_{h}}{2} \right)} - \coth{\left(\frac{\beta_{c} \omega_{c}}{2} \right)} \right\},
\label{spectrumJJ}
\end{equation}
for gravitational, electromagnetic, and conformal-coupling scalar field (figure \ref{fig:spectrumM} and \ref{fig:spectrumJ}).
The scalar field significantly contributes to the mass-loss rate of the black hole at lower spin parameters. On the other hand, the energy flux of gravitons is dominant at higher-spin parameters due to the superradiance. In fact, the rate of change of mass associated with the scalar field of $(l,m_{z}) = (0,0)$, which does not superradiate, is \textit{positive} for the near-extremal case where the black hole temperature is smaller than that of the cosmological horizon. 
However, the contribution of this mode is negligible compared to the other components, and the total rate of change of mass is still negative.
The mass-loss and spin-loss rates, $dM/dt$ and $dJ/dt$, are computed in the spin-mass parameter space in figure \ref{fig:flux_global} for graviton-emission as an example. (The computation of $M(t)$ and $J(t)$ including the contribution of multiple species and modes are performed and the results are shown in FIG. \ref{fig:time_development} and \ref{fig:separated}.). For the Nariai case, $r_{h} =r_{c}$, the horizon frequencies, $\Omega_{h}$ and $\Omega_{c}$, and horizon temperatures, $1/\beta_{h}$ and $1/\beta_{c}$, are equivalent (figure \ref{fig:ratio_temp}). Nevertheless, the energy flux from the black hole does not balance exactly with that from the cosmological horizon. The energy (angular momentum) flux $dM/dt$ ($dJ/dt$), measured in the Boyer-Lindquist time $t$, is suppressed by $(r_{c}-r_{h})$ near the Nariai limit (see appendix \ref{app:Nariai}) whereas the blueshift factor, $dt/d\tau \propto 1/\sqrt{\Delta_{r}}$, is proportional to $(r_{c}-r_{h})^{-1}$, where $\tau$ is the proper time of an observer in the region between the two horizons. Therefore, the energy and angular momentum fluxes measured by the observer, $dM/d\tau$ and $dJ/d\tau$, are in general finite and non-zero, even in the Nariai limit.

\begin{figure} 
\centering
\includegraphics[width=0.7\linewidth]{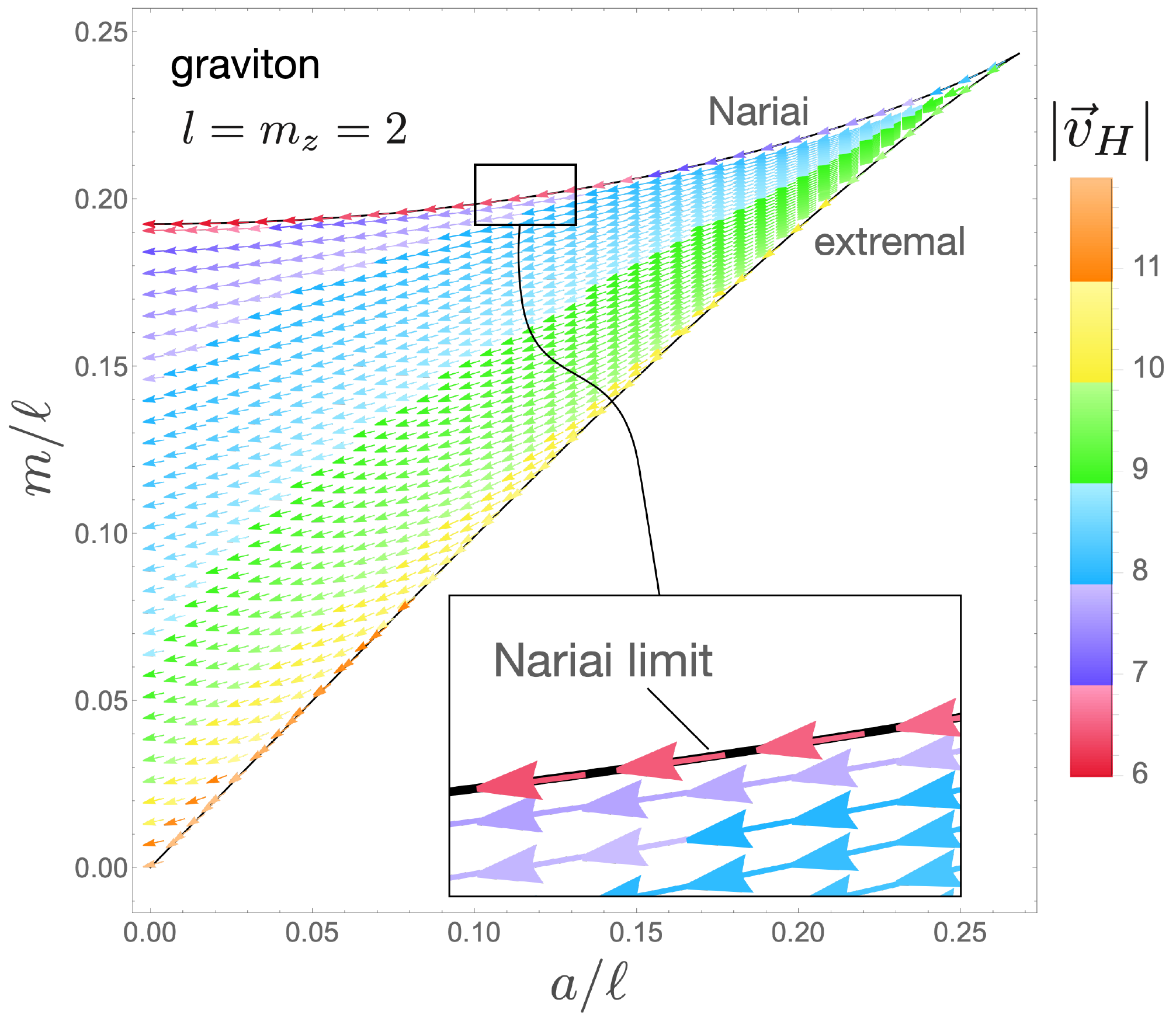}\hspace{1cm}
\caption{Stream map showing the time evolution of the KdS spacetime by taking into account graviton emission only ($l=m_{z} = 2$).
} \label{fig:flow_gra}
\end{figure}

\begin{figure} 
\centering
\includegraphics[width=0.7\linewidth]{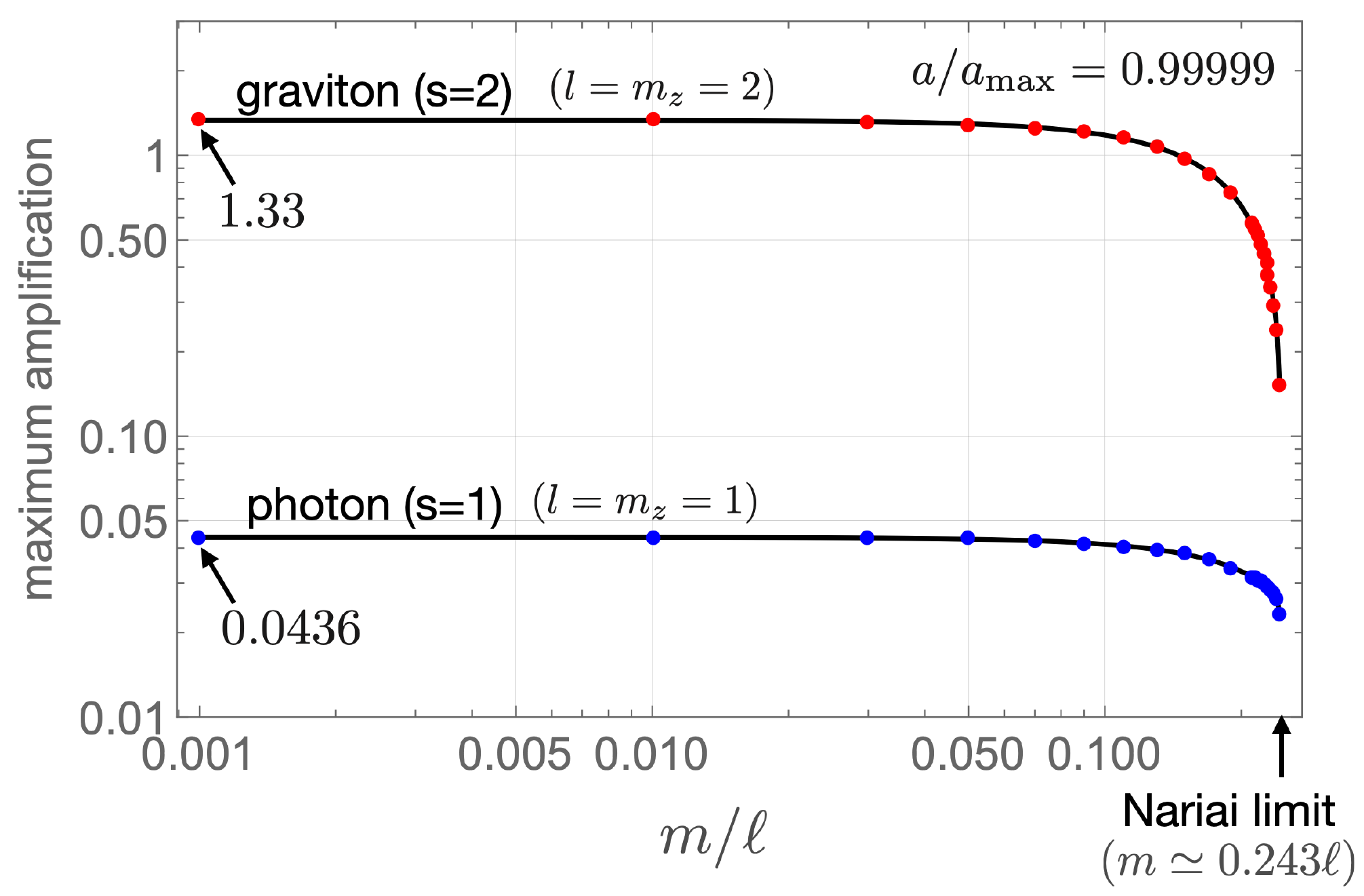}\hspace{1cm}
\caption{The maximum values of the greybody factors of graviton ($l=m_{z} =2$) and photon ($l=m_{z} =1$) for various values of $m$. We take the near-extremal limit with $a/a_{\rm max} = 0.99999$.
} \label{fig:ampl}
\end{figure}

We evaluate the vector field
\begin{equation} \label{flow}
\vec{v}_{H} (a,m) \equiv \left(\frac{da}{dt}, \frac{dm}{dt} \right),
\end{equation}
and it is shown in figure \ref{fig:flow_gra}.
One might wonder whether there are any trajectories which end at the Nariai limit, since this would be a counter example to the expectation that black holes should eventually completely evaporate away. However, our calculations indicate that the flow in \eqref{flow} runs parallel to the Nariai line, which means that the Nariai solution cannot be an attractor solution\footnote{The extremal or (non-Nariai) equal-temperature limits cannot be an attractor since the energy flux is non-zero in those limits. In order for the net flux to be zero, $\beta_{h} = \beta_{c}$ and $\Omega_{h} = \Omega_{c}$ should be satisfied.}, at least when graviton-emission dominates the Hawking radiation.
As seen from figure \ref{fig:flux_global} and \ref{fig:flow_gra}, energy and angular momentum fluxes are higher near the extremal and lower-mass region. This is relevant to the intensity of superradiance which can be quantified by the greybody factor. We evaluated the maximum value of the greybody factor for $a/a_{\rm max} = 0.99999$ as a function of $m$ (see figure \ref{fig:ampl}). We find out that the cosmological constant quenches the effect of superradiance. One can see that the value of amplification factor in the limit of $m/\ell \to 0$ goes to the maximum amplification factor for Kerr spacetime, which was evaluated in \cite{1974ApJ...193..443T,Rosa:2016bli}.

\subsection{Time evolution}

\begin{figure} [h]
\centering
\includegraphics[width=1\linewidth]{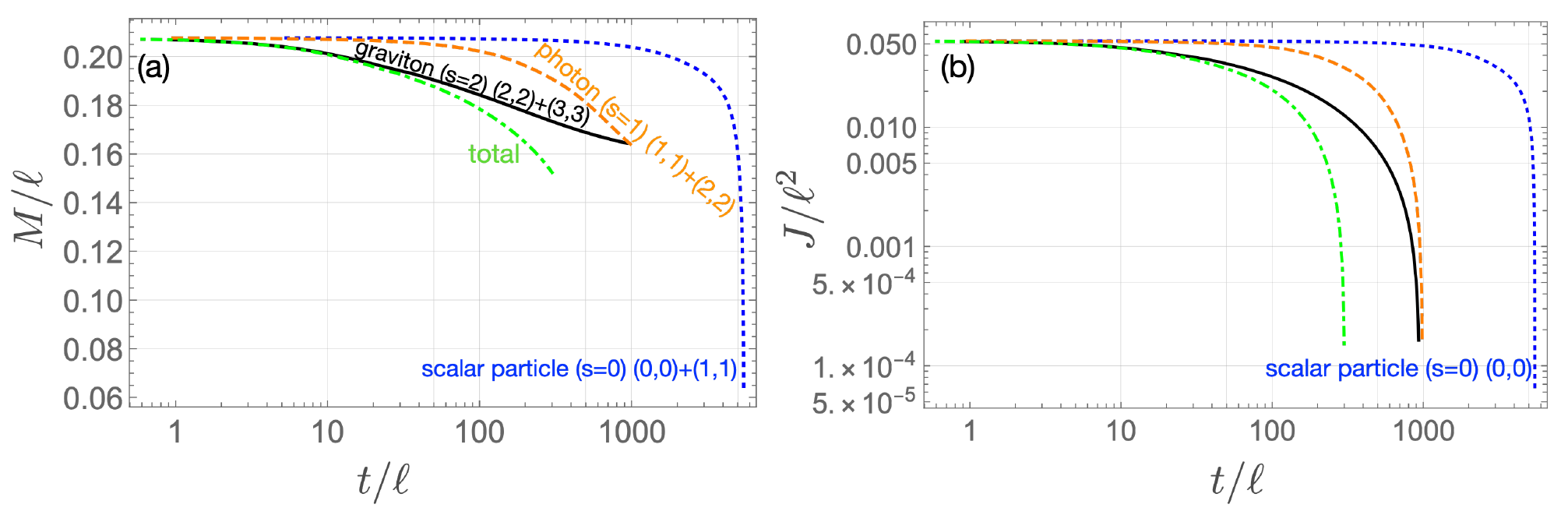}\hspace{1cm}
\caption{The time-development of (a) mass and (b) angular momentum with a near-extremal initial condition of $(a,m) =(0.255 \ell, 0.2355 \ell)$.} \label{fig:time_development}
\end{figure}

We compute the time evolution of a spinning black hole in the near-extremal limit by taking into account gravitons with $(l,m_{z}) = (2,2)$ and $(3,3)$, photons with $(1,1)$ and $(2,2)$, and scalar perturbations with $(0,0)$ and $(1,1)$\footnote{We checked that other higher modes contribute less than $5 \%$ compared to the modes we take into account, at least when $m=0.1$ and $a/a_{{\rm max}}\geq 0.1$.}. We numerically integrate the differential equations (\ref{z_evo}) and (\ref{tau_evo}) by the 4th-order Runge-Kutta algorithm at 40 steps in $y$. The result is presented in figure \ref{fig:time_development} where we show the time-development of $J$ and $M$ computed individually for each component. One can see that the spin and mass-loss rates are dominantly governed by graviton-emission. This is consistent with the spectra shown in figure \ref{fig:spectrumM} and \ref{fig:spectrumJ}.
\begin{figure} [h]
\includegraphics[width=1\linewidth]{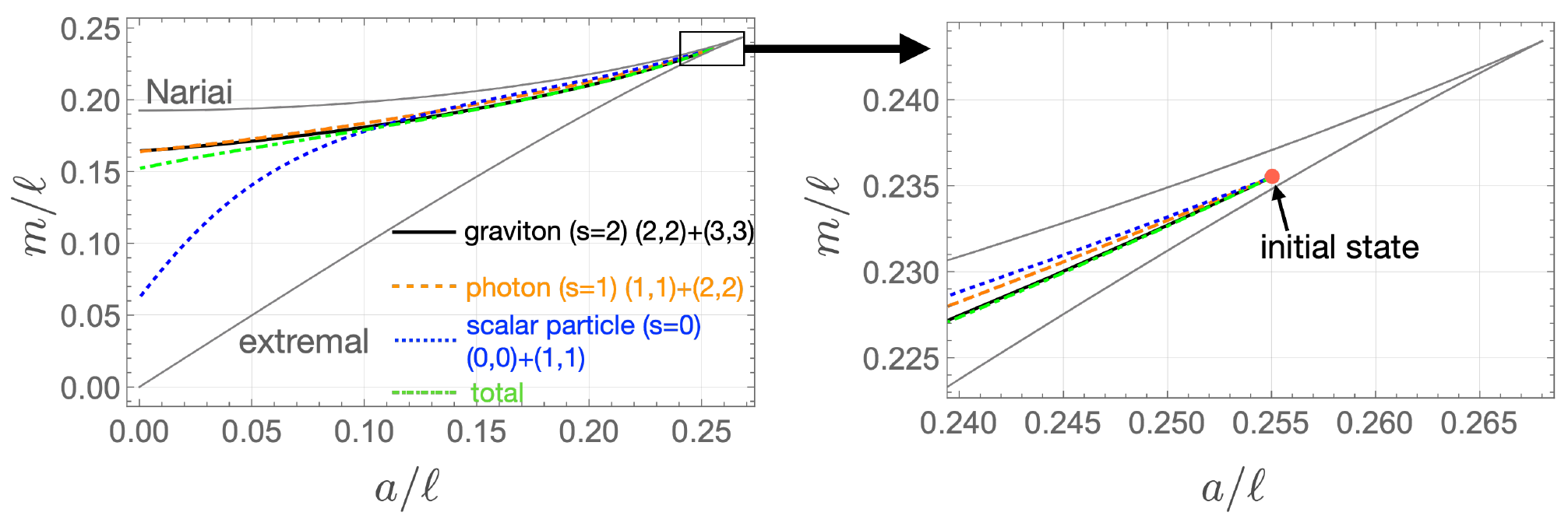}\hspace{1cm}
\caption{Trajectories of the KdS black hole on the $a$-$m$ plane. We set the same initial condition as in figure \ref{fig:time_development}.
} \label{fig:separated}
\end{figure}
The trajectories of the KdS black hole are shown in the spin-mass parameter space (figure \ref{fig:separated}). In the slow-rotation case, the $(0,0)$ component of the scalar emission (monopole radiation) carries out the large amount of mass energy since a slowly spinning black hole is almost spherical.

\begin{figure} [h]
\centering
\includegraphics[width=1\linewidth]{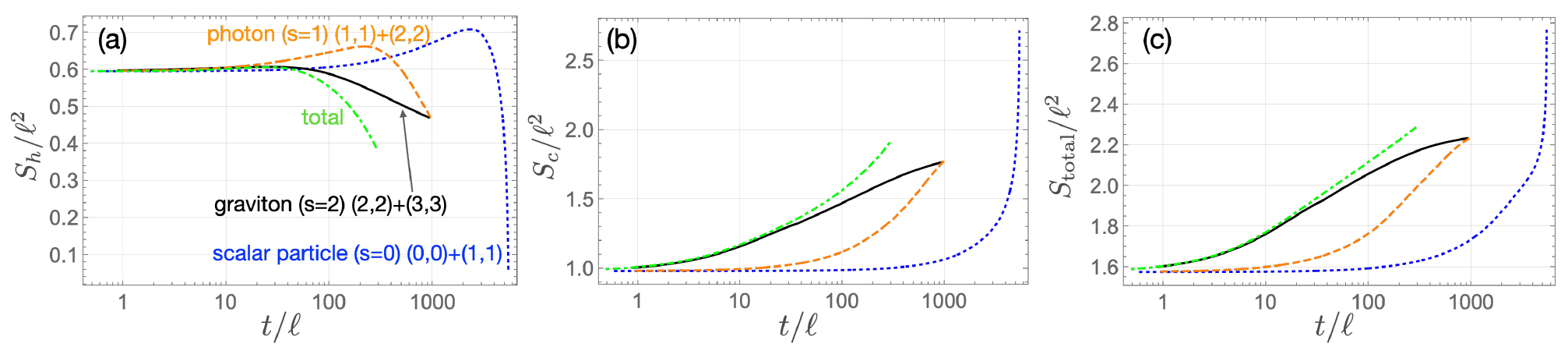}\hspace{1cm}
\caption{The time-development of the Bekenstein-Hawking entropies, $S_{\rm h}$, $S_{\rm c}$, and $S_{\rm total}$, are shown. We set the same initial condition as in figure \ref{fig:time_development}.
} \label{fig:entropy_time}
\end{figure}

The thermodynamics of the 
black hole and cosmological horizons have been investigated by
Dolan et al.\ \cite{Dolan:2013ft}.
It is interesting to consider the evolution of the entropy of the KdS black hole system,
as typically the black hole horizon will shrink, however, the cosmological horizon also
grows as the black hole loses mass and spins down. In figure \ref{fig:entropy_time}, we show the time-development of the Bekenstein-Hawking entropy of the KdS black hole and cosmological horizon, which have the forms of
\begin{equation}
S_{h} \equiv \frac{\pi (r_h^2+a^2)}{\Xi}, \qquad S_{c} \equiv \frac{\pi (r_c^2+a^2)}{\Xi},
\end{equation}
respectively, and the total entropy is $S_{\rm total} \equiv S_{h} + S_{c}$. As conjectured by the generalized second law of thermodynamics, we confirm that the total entropy monotonically increases, which means that the generalized second law is satisfied even for the evaporation process of a KdS black hole.

\section{Conclusion}
In this paper, we computed the time evolution of a KdS black hole formed from 
gravitational collapse in de Sitter space. We argued for a particular choice of 
vacuum state that was consistent with the absence of the past cosmological 
horizon, and applicable to the inflationary universe scenario. The usual Bunch-Davies vacuum state
is transformed by collapse into an excitation of an Unruh-like vacuum state. The Hawking
stress-energy flux causes the black hole mass and rotation to decay, but has no effect on the 
cosmological constant.

We derived the spectrum of the energy flux based on these assumptions and computed the greybody factor
of the radiation using analytic solutions of the Teukolsky equations on the KdS background. Note that the 
Teukolsky equations for a massless scalar field reduce to Heun's equation, but only when it is conformally 
coupled to gravity. For this reason we have taken the massless scalar field in our analysis to be conformally 
coupled to gravity.  The scalar field  (which has a zero angular momentum mode) contributes 
most of the mass-loss rate at lower black hole spins, while the graviton significantly contributes to mass-loss and 
spin-loss rates for a rapidly spinning black hole. The rapidly spinning case is dominated
by superradiance around the KdS black hole, but the superradiance is weakened by a large value of cosmological constant.
We have computed numerically the time evolution of a KdS black hole taking into account dominant modes of scalar, electromagnetic, 
and gravitational perturbations. In this paper, for simplicity we considered 
bosonic particles only, leaving fermions for future work.
We found, like in the asymptotically flat case, that a rotating black hole in de Sitter space loses angular momentum faster than it loses mass. We also argued that there are no black holes which evolve toward the Nariai limit.

We have also computed the horizon entropy and have found that the generalized 
second law of thermodynamics still holds thanks to the 
cosmological constant, at least within the parameter region we have investigated numerically.
Knowledge of the heat fluxes raises the possibility of combining this
with concepts from non-equilibrium thermodynamics. For example, at a linearized level we might take interacting fields
and study the transport properties of the medium between the horizons, taking into account heat flows
and entropy production for the interacting fluxes. It would also be interesting to
explore the similarities to those systems with evolving heat fluxes in extended irreversible thermodynamics \cite{Jou_1988}.

\acknowledgments

This work was supported in part by the Leverhulme Trust 
[Grant No. RPG-2016-233] (RG/IGM/SP), by the
STFC [Consolidated Grant ST/T000708/1] (RG/IGM), 
by JSPS Overseas Research Fellowships (NO), by the Perimeter Institute for 
Theoretical Physics (RG/NO) and by the Natural Science and Engineering Research Council of Canada (SP). Research at Perimeter Institute is supported by 
the Government of Canada through the Department of Innovation, 
Science and Economic Development Canada and by the Province of 
Ontario through the Ministry of Research, Innovation and Science.

\appendix

\section{Normalisation}
\label{app:Normalization}

We will explain here the normalisation of the modes given in Eq.\ (\ref{modes}) 
for the expansion (\ref{expand}).
The norm of a state is defined by \cite{PhysRevD.10.3194},
\begin{equation}
\langle \phi,\phi\rangle = i\int_\Sigma\sqrt{-|g|}g^{t\mu}(\phi\,\phi^*_{,\mu}-\phi_{,\mu}\phi^*)d^3x,
\end{equation}
where $\Sigma$ is a constant time surface.  The various contribution are,
\begin{equation}
\sqrt{-|g|}={\rho^2\sin\theta\over\Xi},
\end{equation}
and close to the horizons,
\begin{equation}
g^{t\mu}\partial_\mu\to-{(a^2+r^2)^2\over \Delta_r\rho}\hat\partial_t
\end{equation}
where $\hat\partial_t=\partial_t-\Omega\partial_\varphi$. Finally,
\begin{equation}
drd\theta d\varphi={\Delta_r\over a^2+r^2}dr^*d\theta d\varphi
\end{equation}
Hence, near the horizons,
\begin{equation}
\langle \phi,\phi\rangle = -i\int{a^2+r^2\over \Xi}(\phi\hat\partial_t\phi^*-\hat\partial_t\phi\,\phi^*)\sin\theta\,dr^*d\theta d\varphi.
\end{equation}
To normalise, we insert the expansion of $\phi$ for two generic modes,
\begin{equation}
\begin{split}
\langle \phi_{\omega l m_z},\phi_{\omega' l' m_z'}\rangle = & \ \frac{2\pi}{\Xi}\delta_{m_z m_z'}(\omega_{c,h}+\omega'_{c,h})e^{-i(\omega-\omega')t} \\
& \ \ \times \int d\theta\sin\theta S_{\omega l m_z}S_{\omega' l' m_z'} \int dr_*R_{\omega l m_z}R^*_{\omega' l' m_z'}.
\end{split}
\end{equation}
By constructing wave packets in the region containing the incident part of the wave, we can substitute ${\overleftarrow R}_{\omega l m_z}=\mathcal{N}_c e^{-i\omega_c r_*}$ at $r=r_c$ and ${\overrightarrow R}_{\omega l m_z}=\mathcal{N}_h e^{i\omega_h r_*}$ at $r=r_h$, where $\mathcal{N}_{c,h}$ are the normalisation factors to be found.
The integral in $r_*$ then becomes $|\mathcal{N}_{c,h}|^2\delta_{\omega\omega'}$ and the angular functions are normalised such that the $\theta$ integral becomes $(2\pi)^{-1}$ for the same $l$ mode.
Thus, we find $\mathcal{N}_{c,h}=\sqrt{\Xi/|2\omega_{c,h}|}$ as claimed in \eqref{modes}.

\section{Blueshifted flux in the Nariai limit and its finiteness}
\label{app:Nariai}

We investigate the behaviour of the spectra $d^{2} M /dt/d\omega$ and $d^{2} J /dt/d\omega$ in the Nariai limit and show that the flux measured by an observer in the region between the black hole and cosmological horizon is still finite value even in the Nariai limit. The spectra shown in (\ref{spectrumMM}) and (\ref{spectrumJJ}) can be approximated as follow in the Nariai limit and $\omega < m_{z}\Omega_{c} (\omega_{c} < 0)$:
\begin{align}
\begin{split}
\frac{d^{2} M}{dt d\omega} &= \frac{1}{2 \pi} (\omega + am_{z}/\ell^{2}) {}_{s} \Gamma_{\omega l m_{z}} \frac{1}{2} \left\{ \coth{\left( \frac{\beta_{h} \omega_{h}}{2} \right)} - \coth{\left( \frac{\beta_{c} \omega_{c}}{2} \right)}  \right\} \\
&= \frac{1}{4 \pi} (\omega + am_{z}/\ell^{2}) {}_{s} \Gamma_{\omega l m_{z}} \left( - \frac{1+\exp(\beta_{h} \omega_{h})}{1-\exp(\beta_{h} \omega_{h})} + \frac{1+\exp(\beta_{c} \omega_{c})}{1-\exp(\beta_{c} \omega_{c})} \right)\\
&\simeq \frac{1}{2 \pi} (\omega + am_{z}/\ell^{2}) {}_{s} \Gamma_{\omega l m_{z}} \left( \exp(\beta_{c} \omega_c) - \exp(\beta_{h} \omega_h) \right) \to 0,
\end{split}
\end{align}
where we used $\beta_{h}, \beta_{c} \to \infty$ in the Nariai limit. From this estimation, one can conclude that $d^{2} J/dt/d\omega$ is also exponentially suppressed in $\omega < m_{z}\Omega_{c}$. When $\omega > m_{z}\Omega_{h}$, we have
\begin{align}
\begin{split}
\frac{d^{2} M}{dt d\omega} &= \frac{1}{2 \pi} (\omega + am_{z}/\ell^{2}) {}_{s} \Gamma_{\omega l m_{z}} \frac{1}{2} \left\{ \coth{\left( \frac{\beta_{h} \omega_{h}}{2} \right)} - \coth{\left( \frac{\beta_{c} \omega_{c}}{2} \right)}  \right\} \\
&= \frac{1}{4 \pi} (\omega + am_{z}/\ell^{2}) {}_{s} \Gamma_{\omega l m_{z}} \left(\frac{1+\exp(-\beta_{h} \omega_{h})}{1-\exp(-\beta_{h} \omega_{h})} - \frac{1+\exp(-\beta_{c} \omega_{c})}{1-\exp(-\beta_{c} \omega_{c})} \right)\\
&\simeq \frac{1}{2 \pi} (\omega + am_{z}/\ell^{2}) {}_{s} \Gamma_{\omega l m_{z}} \left( \exp(-\beta_{h} \omega_h) - \exp(-\beta_{c} \omega_c) \right) \to 0,
\end{split}
\end{align}
from which we also have $d^{2}J/dt/d\omega \to 0$ for $\omega > m_{z}\Omega_{h}$ in the Nariai limit. 
\begin{figure} [t]
\centering
\includegraphics[width=0.7\linewidth]{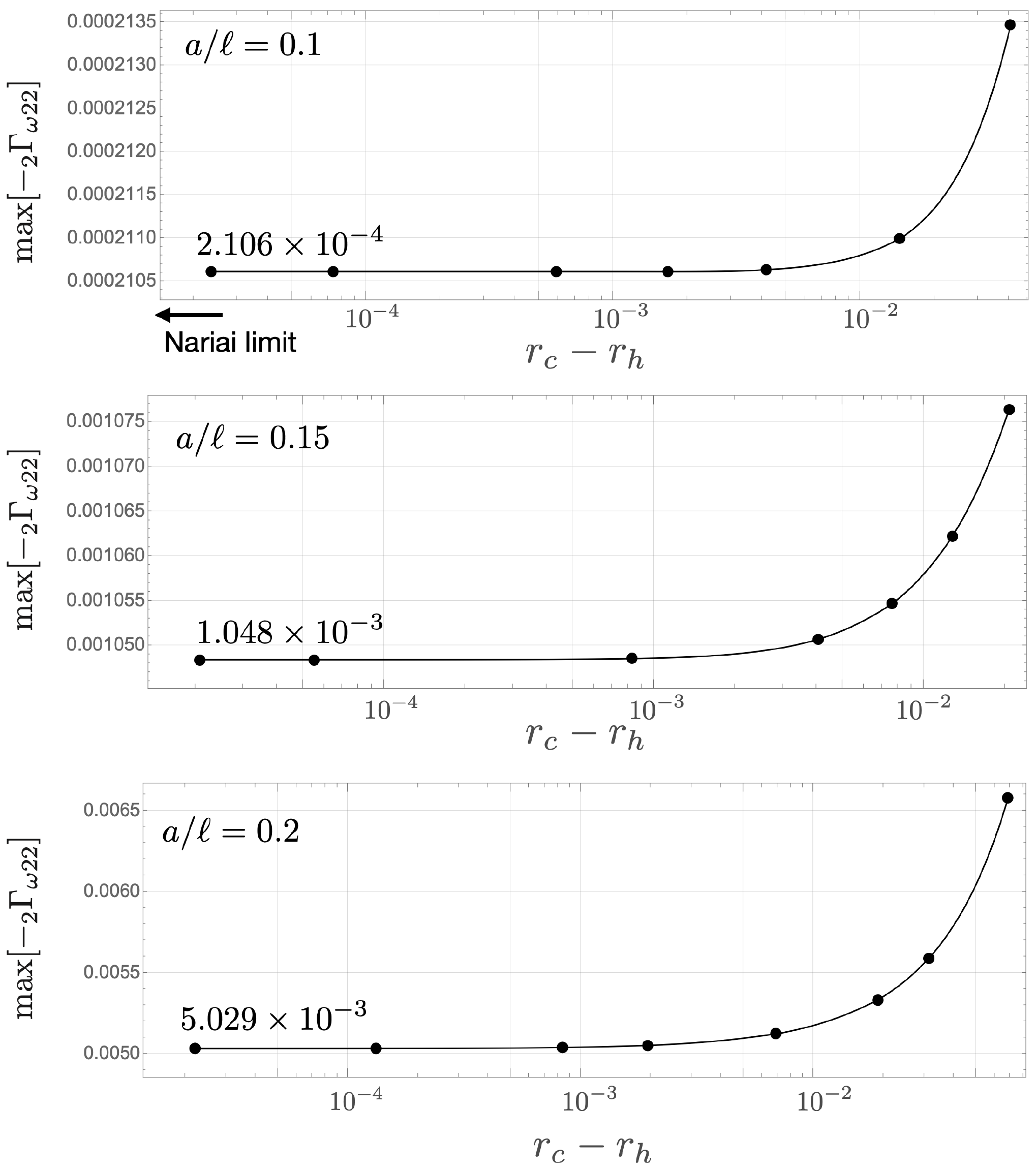}\hspace{1cm}
\caption{The maximum values of the greybody factor for graviton ($s=2$) with $(l,m_z) = (2,2)$ near the Nariai solution for $a/\ell=0.1$, $0.15$, and $0.2$.
} \label{fig:nariai_grey}
\end{figure}
Let us next consider the superradiant frequencies, $m_{z}\Omega_{c} < \omega < m_{z}\Omega_{h}$, in the Nariai limit. The spectrum is
\begin{align}
\begin{split}
\frac{d^{2} M}{dt d\omega} &= \frac{1}{2 \pi} (\omega + am_{z}/\ell^{2}) {}_{s} \Gamma_{\omega l m_{z}} \frac{1}{2} \left\{ \coth{\left( \frac{\beta_{h} \omega_{h}}{2} \right)} - \coth{\left( \frac{\beta_{c} \omega_{c}}{2} \right)}  \right\}, \\
&= \frac{1}{4 \pi} (\omega + am_{z}/\ell^{2}) {}_{s} \Gamma_{\omega l m_{z}} \left( \frac{2}{\beta_{h} \omega_{h}} - \frac{2}{\beta_{c} \omega_{c}} \right), \\
&\simeq -\frac{1}{2\pi} (\omega + am_{z}/\ell^{2}) {}_{s} \Gamma_{\omega l m_{z}} \beta_{h}^{-1} \frac{1}{k (1-k)} \frac{1}{\Delta \omega} \propto (r_{c}-r_{h})^{0},
\end{split}
\label{spectrumMoh}
\end{align}
where $\Delta \omega \equiv m_{z} (\Omega_{h} - \Omega_{c})$ and we set $\omega = m_{z}\Omega_{c} + k \Delta \omega$ with $0 < k < 1$. Since both $\beta_{h}^{-1}$ and $\Delta \omega$ are proportional to $(r_{c}-r_{h})$, the spectrum amplitude is still finite and non-zero even in the Nariai limit for the superradiant frequencies. Then we found out that $dM/dt$ and $dJ/dt$ are proportional to $\Delta \omega \propto r_{c}-r_{h}$.

An observer in the region between the horizons measures $dM/d\tau = (dt/d\tau) dM/dt$ and $dJ/ d\tau =(dt/d\tau) dJ/dt$, where $\tau$ is the proper time of the observer. The factor $dt/d\tau$ has the blueshift factor $1/\sqrt{\Delta_{r}}$ which enhances the intensity of flux by $\sim 1/(r_{c}-r_{h})$ in the Nariai limit. Therefore, the gravitational blueshift cancels the suppression of $(r_{c}-r_{h})$ in $dM/dt$ and $dJ/dt$, and in general, the measured fluxes $dM/d \tau$ and $dJ/ d \tau$ are still finite and non-zero, even in the Nariai limit.

Note that $1/k(k-1)$ in (\ref{spectrumMoh}) does not lead to the divergence of $d^{2}M/dt/d\omega$ at $\omega \to m_{z} \Omega_{c/h}$ because the greybody factor ${}_{s} \Gamma_{\omega l m_{z}}$ suppresses the flux with the power of $(\omega - m_{z}\Omega_{c/h})$. Also, the greybody factor has its positive value even in the Nariai limit for superradiant frequencies and the maximum values of greybody factor for fixed spin parameters are shown in figure \ref{fig:nariai_grey}.

Finally, let us evaluate $dM/d\tau$ in the non-spinning Nariai limit. When $a=0$, the energy flux is
\begin{align}
\begin{split}
\frac{d M}{dt} &= \int_{0}^{\infty} d\omega \frac{\omega}{2 \pi} {}_{s} \Gamma_{\omega l m_{z}} \frac{1}{2} \left\{ \coth{\left( \frac{\beta_{h} \omega}{2} \right)} - \coth{\left( \frac{\beta_{c} \omega}{2} \right)}  \right\}\\
&\simeq \int_{0}^{\infty} d\omega \frac{\omega}{2 \pi} {}_{s} \Gamma_{\omega l m_{z}} \left( \exp(-\beta_{h} \omega) - \exp(-\beta_{c} \omega) \right)\\
&\sim \frac{{}_{s} \Gamma_{0 l m_{z}}}{2 \pi} \int_{0}^{\infty} d\omega \omega \left( \exp(-\beta_{h} \omega) - \exp(-\beta_{c} \omega) \right)\\
&= \frac{{}_{s} \Gamma_{0 l m_{z}}}{2 \pi} \left( \frac{1}{\beta_{h}^{2}} - \frac{1}{\beta_{c}^{2}} \right)\\
&\simeq \frac{{}_{s} \Gamma_{0 l m_{z}}}{2 \pi} \frac{2}{\beta_{h}} \left( \frac{1}{\beta_{h}} - \frac{1}{\beta_{c}} \right) \propto (r_{c}-r_{h})^{3},
\end{split}
\label{non_rotSPE}
\end{align}
where we assumed a special situation where $\displaystyle \lim_{\omega \to 0} |{}_{s} \Gamma_{\omega l m_{z}}| >0$ \cite{Kanti:2005ja,Kanti:2014dxa} although the greybody factor is suppressed by a power of $\omega$ in most cases, which increases the power of $(r_{c} -r_{h})$ in the last line of (\ref{non_rotSPE}).
Therefore, taking into account the blueshift factor $\sim 1/(r_{c}-r_{h})$, we have $dM/d\tau \propto (r_{c}-r_{h})^{2}$ which means that it goes to zero in the Nariai limit.

\end{document}